\documentclass[12pt]{article}
\usepackage{graphicx}
\usepackage{amssymb}
\newcommand{\Z}{{\mathbb Z}}
\title{The non-abelian Born-Infeld action at order $F^6$}
\author{A. Sevrin, J. Troost and W. Troost}
\begin{document}
\thispagestyle{empty}
\begin{minipage}[t]{5cm}
\small
{\tt hep-th/0101192}\\ 
January 26, 2001
\normalsize
\end{minipage}
\hspace{5cm}
\begin{minipage}[t]{4cm}
\begin{flushright}
\small
VUB/TENA/01/01\\
MIT-CTP-3074\\
KUL-TF-2001/02
\normalsize
\end{flushright}
\end{minipage}
\addtocounter{footnote}{1}

\vspace{1cm}

\begin{center}
{\bf \large The non-abelian Born-Infeld action at order $F^6$}

\vspace{1.8cm}

Alexander Sevrin${}^1$, Jan Troost${}^2$
and Walter Troost${}^3$
\vspace{.2cm}

${}^1${\em Theoretische Natuurkunde, Vrije Universiteit Brussel} \\
{\em Pleinlaan 2, B-1050 Brussels, Belgium} 

\vspace{.1cm}

${}^2${\em Center for Theoretical Physics, MIT,}\\
{\em 77 Mass Ave, Cambridge, MA 02139, USA}

\vspace{.1cm}

${}^3${\em Instituut voor theoretische fysica, Katholieke  Universiteit Leuven,}\\
{\em Celestijnenlaan 200D, B-3001 Leuven, Belgium} 

\end{center}


\small
\abstract{To gain insight into the non-abelian Born-Infeld (NBI) action,
we study coinciding D-branes wrapped on tori, and turn on magnetic fields on
their
worldvolume. We then compare predictions for the spectrum of open strings
stretching between these D-branes,
from perturbative string theory and from the effective NBI action. Under
some plausible assumptions, we find corrections to the Str-prescription
for the NBI action at order $F^6$. In the process we give a way to classify
terms in the NBI action that can be written in terms of field strengths only,
in terms of permutation group theory.}
\noindent



\section{Introduction}
Consider a flat D-brane in type II string theory. The bosonic
massless degrees of
freedom of an open string ending on the D-brane are a $U(1)$ gauge field,
associated to excitations of the string longitudinal to the brane, and
neutral scalar fields, associated to transverse excitations of the brane.
The effective action for these massless degrees of freedom for slowly
varying field strengths is known up to all orders in the string length
$\sqrt{\alpha'}$. It is the Born-Infeld action
\footnote{The Dp-brane tension we denote $T_p$, and
$\alpha \in \{0,1,\dots,p\}$.
We choose the static gauge  and
we leave out the transverse scalars for reasons to be explained below.}:
\begin{eqnarray}
{\cal S} &=& - T_p \int d^{p+1} \sigma \,
\sqrt{\det\left(\delta_{\alpha}{}^{\beta} + 2 \pi \alpha' 
F_{\alpha}{}^\beta\right)}.
\end{eqnarray}
Expanding this action in the field strength, we
obtain a Maxwell action with higher order corrections in $\alpha' F$.

When $N$ D-branes coincide, the massless degrees of freedom of open
strings beginning and ending on them are a $U(N)$ gauge field, and a
number of
scalar fields in the adjoint of the gauge group. The extra degrees
of freedom come from strings stretching from one D-brane to another that
become massless
when these D-branes coincide. A problem that seems to
appear naturally by analogy with the abelian case,
is to write down the effective action for these massless degrees of
freedom, for slowly varying field strengths. In fact, we know the
first terms of such a non-abelian Born-Infeld (NBI) action exactly. From string
scattering amplitudes \cite{T} and a three-loop betafunction
calculation \cite{BP}, we know
that the expansion of the NBI lagrangian  in powers of the field strengths
begins with
\footnote{We put
$2 \pi \alpha'=1$ from now on, ignore the overall factor $T_p$, and an additive constant.}:
\begin{eqnarray}
{\cal L} &=& Tr ( \frac{1}{4} F^{\alpha_1 \alpha_2} F_{\alpha_2 \alpha_1} \nonumber \\
& & + \frac{1}{24} F^{\alpha_1 \alpha_2} F_{\alpha_2 \alpha_3} F^{\alpha_3 \alpha_4}
F_{\alpha_4
\alpha_1} + \frac{1}{12} F^{\alpha_1 \alpha_2} F^{\alpha_3 \alpha_4} F_{\alpha_2 \alpha_3}
F_{\alpha_4 \alpha_1} \nonumber \\
& & - \frac{1}{48} F^{\alpha_1 \alpha_2} F_{\alpha_2 \alpha_1}
                   F^{\beta_1 \beta_2} F_{\beta_2 \beta_1}
   -  \frac{1}{96} F^{\alpha_1 \alpha_2} F^{\beta_1 \beta_2}
                   F_{\alpha_2 \alpha_1} F_{\beta_2 \beta_1}    \nonumber \\
& & + {\cal O}(F^6) ).  \label{act4}
\end{eqnarray}
Through this order, this coincides with the expansion of the symmetrized
trace action \cite{Tstr}:
\begin{eqnarray}
{\cal S} &=& -\int d^{p+1} \sigma \, Str (\det\sqrt{\delta_{\alpha}{}^{\beta}+
 F_{\alpha}{}^{\beta}} ),
\end{eqnarray}
where the prescription is to formally expand the square root and the determinant
in $F$ first,  then to symmetrize over
 all orderings of the field strength factors,
and finally to perform the trace.

\pagebreak[2]
\label{discussion1}
There is some ambiguity in the expression of the NBI action in terms of field
strengths and their covariant derivatives, since
$[D_{\alpha}, D_{\beta}] F_{\gamma \delta} = i [F_{\alpha \beta}, F_{\gamma
\delta}]$. One could rewrite expression (\ref{act4}) by assembling the second
with the third term (and the fourth with the fifth) at the cost of introducing
extra $[D,D] F F F$ terms. The all order proof \cite{Tstr} of the symmetric trace formula
is only claimed to be valid up to this type of terms,
and therefore pertains only to the sum of the
coefficients of the second and third terms (and likewise for the fourth and
fifth), which in fact also follows from specialising to the abelian case.
Nevertheless it is remarkable that, to fourth order, the symmetric trace gives
the complete expression for the superstring (though not for the bosonic string),
and thus it deserves to be investigated in detail at higher
orders.
\label{discussion2}
In this paper, we will embed the symmetric trace hypothesis into a more general
action. Since we are approximating the NBI at string tree level,
we do keep the restriction of considering only
an overall trace in the fundamental over the gauge group factors.
Expanding for `slowly varying field strengths' is admittedly ambiguous,
and an unambiguous order would add the number of  $ F$'s to twice the number of $D$'s.
We will not include the most general possibility, but limit
ourselves to a subset adapted to the exploratory program that we propose
in the next section: all terms where the covariant derivatives occur in
antisymmetric combinations, and can therefore be written purely in terms of the
field strengths, are included in our analysis,
but symmetric derivative combinations are not. In other words, we adopt
here the definition that acceleration terms are expressed as symmetrized
products of covariant derivatives.

A direct calculation of the $F^6$ terms would imply the study of a
6-gluon open string amplitude or a 5-loop $\beta$-function. Both are
technically very involved. In the next we will develop a simpler approach
which will allow us to determine the $F^6$ term to a large extent.

\section{Wrapped D-branes and the NBI action}
\label{plan}
\subsection*{Magnetic field strengths on tori}\label{ToriFields}
In this section,
we map out our testing ground for any proposal for the NBI action.
Consider $N$ coinciding D$2n$-branes, wrapped around a $T^{2n}$ torus. Switch
on constant magnetic fields in the Cartan subalgebra (CSA) of the $U(N)$ gauge
group. These correspond to embedded D-branes of lower dimension. Choose
the magnetic fields to be blockdiagonal in the Lorentz indices, for
simplicity. The plan \cite{HT} is now to compare the spectrum for small fluctuations
around this background as predicted by string theory, with the spectrum
predicted by the proposed non-abelian Born-Infeld. Since we only
want to consider the
(originally) massless degrees of freedom of the open string, we decouple
the massive modes by sending $\alpha' \rightarrow 0$. To maintain the
relevance of the non-linear corrections to Yang-Mills theory prescribed by the
NBI, we crank up the magnetic field to keep $\alpha' F$ constant.
\subsection*{Perturbative string theory spectrum}
To write down the spectrum for the low-lying modes predicted by perturbative string
theory, we need some notation. Suppose we restrict to the situation in
which we have only 2 D$2n$-branes \footnote{We will do this throughout this
paper. The rationale is that perturbative string theory as well as the linear
analysis we perform is only sensitive to the interactions between each pair
of D-branes \cite{DST}.}. Since the magnetic background is in a CSA,
we can diagonalize it, and associate a magnetic field strength to each of
the two branes, ${\cal F}^{(1)}_{2i-1,2i}$ and ${\cal F}^{(2)}_{2i-1,2i}$.
We chose the
background to be blockdiagonal in the Lorentz indices.
T-dualizing along the 2, 4, ..., $2n$ directions, we end up with two
D$n$-branes at angles given by:
\begin{equation}
\tan \gamma_i^{(n)} = {\cal F}^{(n)}_{2i-1,2i} . \label{relang0}
\end{equation}
Then the modes of the open string connecting the two
D$n$-branes, which
correspond to the off-diagonal
gauge field modes
in the directions $2k-1, 2k$, $k\in\{1,\cdots,n\}$,
have a spectrum \footnote{The modes of
the scalar fields and the fermions have a similar spectrum, and we do not
expect them to
provide any additional information \cite{JT} \cite{DST}.}:
\begin{eqnarray}
M^2_k &=& \sum_{i=1}^n (2 m_i+1) \epsilon_i \mp 2\epsilon_k,
\label{stringspec}\\
\epsilon_i &=& \gamma_i^{(1)} - \gamma_i^{(2)}. \label{relang}
\end{eqnarray}
The details of how to compute this spectrum can be found in \cite{AC} and
some handy formulas are in \cite{DST}.
\subsection*{Yang-Mills analysis}
To gain some
intuition for how this spectrum comes about and to prepare for the
treatment in the case of the effective action, we take a look at the
Yang-Mills approximation to the problem. Consider then the Yang-Mills
truncation of the non-abelian Born-Infeld action. We can study the same
background as before, and determine the spectrum of the fluctuations
around the background in this approximation. This was done in full
detail in \cite{VB} \cite{JT}. The result is:
\begin{equation}
M^2_k = \sum_{i=1}^n \left\{(2m_i+1)
       ({\cal F}_{2i-1,2i}^{(1)}-{\cal F}_{2i-1,2i}^{(2)})\right\}
    \mp         2({\cal F}_{2k-1,2k}^{(1)}-{\cal F}_{2k-1,2k}^{(2)})\,
    \label{ymspec}
\end{equation}
where for convenience, we chose
${\cal F}_{2i-1,2i}^{(1)} > {\cal F}_{2i-1,2i}^{(2)}$. It is clear that
for small field strengths the string spectrum (\ref{stringspec}) reduces to
the Yang-Mills spectrum (\ref{ymspec}), as expected.

The Yang-Mills
spectrum can be argued for as follows. An endpoint of a string ending on
one of these D-branes behaves as an electric charge in a magnetic field.
The corresponding Landau problem has a harmonic oscillator spectrum with
frequency proportional to the magnetic field. The other endpoint
of the string acts as a particle with the opposite charge.
This makes intuitive the fact that for the global motion of
the string, the difference between the field strengths on the two branes
acts as spacing of the energy levels. The zero-point energy, moreover, can be attributed to
a Zeeman splitting of the energy levels due to the fact that different
combinations of the gauge field in directions $2k-1, 2k$ have spin $\pm1$
under the $SO(2)$ associated to these directions.
\subsection*{String theory as rescaled YM}
String theory adds
a non-linearity to this spectrum that can for instance be intuitively
understood in
the T-dual picture, where magnetic fields are interchanged for rotated
branes. (See \cite{HT} and \cite{DST} for instance.) For our
purposes, the important
observation is that the string spectrum is merely a rescaled
Yang-Mills spectrum. Denoting
\begin{eqnarray}
f_i^0=\frac{1}{2} ({\cal F}_{2i-1,2i}^{(1)}+{\cal F}_{2i-1,2i}^{(2)}),\label{8} \\
f_i^3=\frac{1}{2} ( {\cal F}_{2i-1,2i}^{(1)}-{\cal F}_{2i-1,2i}^{(2)}),
\end{eqnarray}
the spectrum is rescaled by a factor
\begin{eqnarray}    \label{rf}\label{schaalfactor}
\alpha_i^2 \equiv  \frac{\epsilon_{i}}{2 f_i^3}
&=& \frac{\arctan{(\frac{2 f_i^3}{1+(f_i^0)^2- (f_i^3)^2})}}{2 f_i^3}
\end{eqnarray} for field strength fluctuations in directions $2i-1,2i$.

A clearcut question is then, whether a proposal for the NBI action reproduces
this rescaled Yang-Mills spectrum predicted by perturbative string
theory. This was investigated in detail for the Str-prescription in
\cite{DST} (expanding
on the initial explorations in \cite{HT} and \cite{B}). For the simplest case,
on $T^2$, the symmetrized trace prescription yielded a
spectrum with the same structure as the Yang-Mills spectrum, but with
incorrect spacings. The disagreement shows up from third order on,
confirming the veracity of the $F^2$ and $F^4$ terms.
This clearly demonstrates
that the Str-prescription is too crude an approximation to the NBI to
yield the correct mass spectrum on our testing ground. For $T^4$, the
situation remained unclear since the
complete spectrum predicted by the Str-action remained undetermined. For BPS configurations
on $T^4$, the
Str-action reproduces precisely the right spectrum, but for other
settings, it seems highly unlikely that the Str-prescription would lead
to the correct results. On $T^6$, the Str would probably not yield the
right spectrum even for BPS configurations \cite{DST}.

\label{discussion3}
The spectrum as predicted by string theory is a
\em rescaled \em Yang-Mills \em spectrum, \em
compare equations (\ref{stringspec}) and (\ref{ymspec}).
Therefore, we will assume that the action relevant for this physical situation,
should yield a \em rescaled \em Yang-Mills \em action \em for the fluctuations,
meaning that it can be brought back to a Yang-Mills action by a suitable
coordinate transformation. This
is certainly the simplest and perhaps the most natural
way to reproduce the desired string theory results.
In these circumstances it seems less natural to allow in the Lagrangian
terms containing derivatives that cannot be written as combinations of field strengths.
Not in the least, they would make it much more difficult in practice to
obtain results for the spectrum, since one would be trying to diagonalize
higher order operators. As indicated before,
to obtain this rescaled Yang-Mills action we do include terms to
the action corresponding to all possible orderings and Lorentz contractions of
field strengths. There might be an a
posteriori justification for this approach,
if one could prove that for the fluctuation eigenfunctions -- they can
explicitly be written down in terms of theta-functions as in \cite{JT} --
other kinds of derivative terms are suppressed.

{From} the formula for the rescaling factor, we expect only terms in the lagrangian
with an even number of field strengths to contribute in our backgrounds.
For this reason, we do not consider terms with an odd number
of field strengths.

\subsection*{BPS conditions}

As already pointed out in \cite{DST}, the translation of the BPS
conditions in string theory in terms of the
background field strength in the effective action  might provide an additional
handle on the NBI action. Concretely, in section \ref{BPS} we will investigate what
constraints on the NBI follow from the demand that self-dual
configurations on $T^4$ should solve the equations of motion.

These constraints on the action are a priori independent from the
ones obtained from the analysis along the line discussed in the previous subsection.
They turn out to provide an independent check on some of the results
obtained with the rescaled YM program, and also to give additional constraints on
the NBI action.

\section{ The NBI at order $F^4$}
\label{order4}
We start by carrying out the program proposed above
at the first non-trivial level, the 
$F^4$ terms in the non-abelian Born-Infeld. This will serve to illustrate 
the method we use in a simple setting. Moreover, it will turn out 
that the straightforward spectral analysis, under the assumptions we make,
is able to replace a four point function computation in open string theory, or
a three-loop beta-function computation in a non-linear $\sigma$-model
approach, demonstrating the power of our method.

The most general lagrangian we can write down under the stated restrictions
(see page \pageref{discussion1}, \pageref{discussion2}, \pageref{discussion3}) is 
then:
\begin{eqnarray}
{\cal L}&=& Tr ( \nonumber \\
& & a^2_1 F_{\alpha_1 \alpha_2} F^{\alpha_2 \alpha_1} \nonumber \\
& & +  a^4_1  F_{\alpha_1 \alpha_2} F^{\alpha_2 \alpha_3} F_{\alpha_3 \alpha_4} F^{\alpha_4 \alpha_1}
+ a^4_2  F_{\alpha_1 \alpha_2} F_{\alpha_3 \alpha_4} F^{\alpha_2 \alpha_3} F^{\alpha_4 \alpha_1} 
\nonumber \\
& & + a^{2,2}_1  F_{\alpha_1 \alpha_2} F^{\alpha_2 \alpha_1} F_{\beta_1 \beta_2} F^{\beta_2 \beta_1}
+ a^{2,2}_2  F_{\alpha_1 \alpha_2} F_{\beta_1 \beta_2} F^{\alpha_2 \alpha_1} F^{\beta_2 \beta_1} 
)\,.
\nonumber \\
& & 
\label{action24}
\end{eqnarray} 
At this low order, it is easy to check that these are indeed the only 
linearly independent terms. At higher order the analysis becomes untransparant.
In section \ref{group}, we will therefore introduce
a diagrammatic representation for these terms. 

The symmetric trace prescription would relate the coefficients 
in equation (\ref{action24}) by $a_2^4=2a_1^4=-4a_1^{2,2}=-8 a_2^{2,2}$
and the determinant formula sets this equal to $\frac{a_1^2}{3}$.
Let us see how this result comes about by imposing the correspondence
of the spectrum with equation(\ref{ymspec}) with the rescaling factor
 (\ref{schaalfactor}).
First of all, we demand that the abelian action be 
reproduced if we restrict to a $U(1)$ subgroup. The 3 constraints this 
yields on the coeficients are easy to determine and they are listed in 
appendix \ref{abelianconstr} equation (\ref{abelian24}).
Next, we determine the action quadratic in off-diagonal gauge field 
fluctuations, in a background blockdiagonal in the Lorentz indices. We
restrict to a $U(2)$ subgroup since we always work with 2 
branes only.  The action for the quadratic fluctuations in this background
is given in appendix \ref{appactquad} equations (\ref{bd2}) and (\ref{bd4}), for 
second and fourth order respectively. Its structure is as follows:
\begin{eqnarray}
{\cal L}^{(2,4)} 
& = & c_i^{\rm kin} (f,a) \left((\delta_1 F_{0, 2i-1}^{(a)})^2 + (\delta_1 F_{0,2i}^{(a)})^2\right) 
  \nonumber \\
& & - c_i(f,a)  (\delta_1 F_{2i-1, 2i}^{(a)})^2
 \nonumber \\
& &  -\frac{1}{2} c_{ij} (f,a) \sum_{i \neq j} \left(
(\delta_1 F_{2i-1, 2j-1}^{(a)})^2 + (\delta_1 F_{2i-1, 2j}^{(a)})^2 + \right.
\nonumber \\ & & \mbox{\hspace*{3cm}}
\left.
 \delta_1 F_{2i, 2j-1}^{(a)})^2   +(\delta_1 F_{2i, 2j}^{(a)})^2 \right)
\nonumber \\ & & 
+ c_{ij}^{\rm nym} (f,a) (\delta_1 F_{2i-1, 2i}^{(a)} \delta_1 F_{2j-1, 2j}^{(a)}) 
 \nonumber \\ & & 
 - 2  c_i^{\rm quad} (f,a) \delta_2 F_{2i-1,2i}^{(3)} f_i^3\,, \label{structure2beYM}
\end{eqnarray}
where (see appendix \ref{notationsappendix}) in $c(.,.)$-coefficients
$f$ represent the background field strength values, 
$a$ stands for the coefficients $a_n^k$ of equation (\ref{action24}),
\begin{eqnarray}
 \delta_1 F_{\alpha \beta} &=& D_{\alpha} \delta A_{\beta}
                              - D_{\beta} \delta A_{\alpha} \,,\\
\delta_2 F_{\alpha \beta} &=&  i [ \delta A_{\alpha}, \delta A_{\beta} ]\,, \\
D_{\alpha} &=& \partial_{\alpha} + i [ A_{\alpha}, . ] \,,
\end{eqnarray}
and the superscript $(a)$ runs over two orthogonal non-CSA $SU(2)$ components.

The different lines are treated as follows:
\begin{itemize}
\item 
The first line is the kinetic term. The first step in the comparison with the 
Yang-Mills action is a rescaling of the fluctuations of the gauge potentials 
such that the kinetic term has the standard normalisation:
 \begin{eqnarray}
     \delta A_n &=& b_i^{-1} \delta a_n \mbox{ for } n\in \{2i-1,2i\}\,,\\
  c_i^{\rm kin} &=& b_i^2\,. \label{herschalingdeltaA}
  \end{eqnarray}
\item The second line represents the deformation energy of the modes
 in directions $2i-1, 2i$. By a  rescaling of the space coordinates,
 \begin{eqnarray}
   X_n &=& b_i \gamma_i x_n \mbox{ for } n\in \{2i-1,2i\}\,,\\
   c_i &=& \gamma_i^2 b_i^4\,,
    \label{herschalingX}
 \end{eqnarray}
it is brought to the standard Yang-Mills form with a rescaled background potential
 \begin{equation}
   a_n = b_i \gamma_i A_n \mbox{ for } n\in \{2i-1,2i\}\,.
    \label{herschalingA}
 \end{equation}
\item In the third line, the rescalings above destroy the Yang-Mills structure 
unless, when $c_{ij} \neq 0$, we have that $\gamma_i=\gamma_j (=\gamma)$.
This being granted, the overall factor agrees with the Yang-Mills value provided  
 \begin{equation}
   c_{ij} = b_i^2 b_j^2 \gamma^2 \,.
 \end{equation}
\item The fourth line is absent from the Yang-Mills action. In accordance with our
assumptions we put their coefficients $c_{ij}^{\rm nym}$ to zero.
\item The fifth line contains the terms linear in the second order fluctuation 
$\delta_2 F_{\mu\nu} = [\delta A_\mu, \delta A_\nu]$.
They have to follow the same scaling as the second line, but in fact this is not
an independent condition. If the Yang-Mills structure of the third line is imposed,
this follows from the fact that fluctuations of the background
configuration that are gauge transformations leave the action unchanged. 
\item Additional terms arise, with the structure
\linebreak[4]
$ \delta_1 F_{2j, 2i}^{(a)} \delta_1 F_{2i-1, 2j-1}^{(a)} -
 \delta_1 F_{2j-1, 2i}^{(a)} \delta_1 F_{2i-1, 2j}^{(a)}\,.$ 
 Partial integration can be combined with the Lie-algebraic structure of these 
 terms to absorb them into the second, fourth and fifth lines.
\end{itemize}

Summarising: the non-Yang-Mills terms have to be put to zero, and then
$ c_i (c_i^{\rm kin})^{-2} =\gamma^2 $
should be independent of $i$, and 
$ c_i (c_i^{\rm kin})^{-1} $
should equal the required scaling factor $\alpha_i^2$ of equation 
(\ref{schaalfactor}).
These demands uniquely fix (after normalizing $a_1^2=1/4$) all coefficients in the action 
(\ref{action24}): 
\begin{eqnarray}
a_1^4&=& \frac{1}{24} \,,\\
a_2^4 &=& \frac{1}{12} \,,\\
a_1^{2,2} &=& -\frac{1}{48}\,, \\
a_2^{2,2} &=& -\frac{1}{96}\,,
\end{eqnarray} 
which matches the action, eq. (\ref{act4}), predicted by the computation of 
scattering amplitudes, a betafunction calculation, and the 
symmetric trace prescription.

\section{Group theory and contractions}
\label{group}
\subsection*{Diagrams} 
The implementation of our program at order 4 in the previous section starts 
from the most general action consisting of terms that
could be written in terms of field strengths alone.
This action is easy to write down in low orders, but at
higher order, a more systematic approach is called for. 
In this section we will describe an attempt to bring some systematics
into  the classification of the different terms
at order 2,4, and 6  by using permutation group theory. 
The explicit examples will be taken from  order 6, and we will also give results at
order 4, but the scheme carries over to  all orders.

\begin{figure}
\setlength{\unitlength}{1mm}
\begin{picture}(120,45)
\put(20,45){\includegraphics[angle=-90,scale=.5]{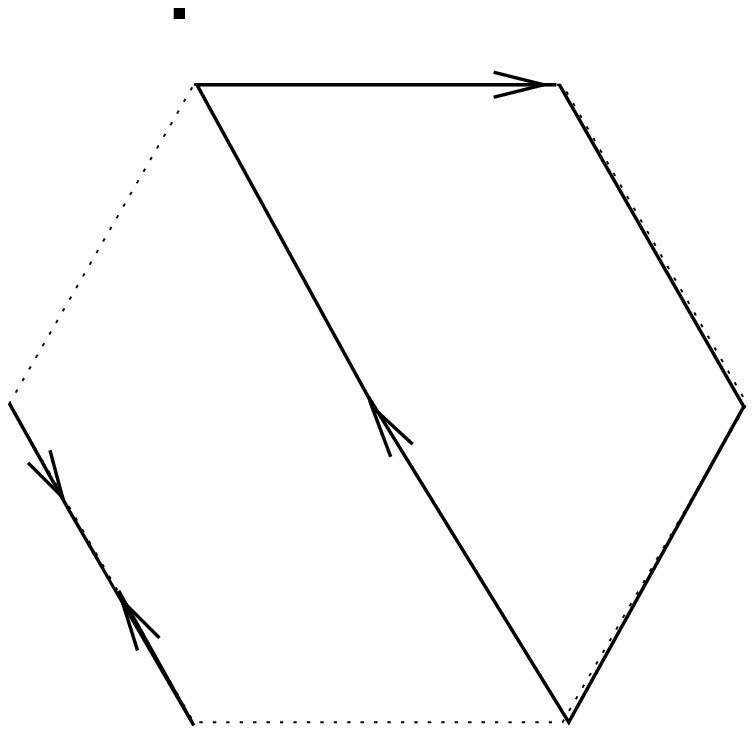}}                                                             
\put(80,45){\includegraphics[angle=-90,scale=.5]{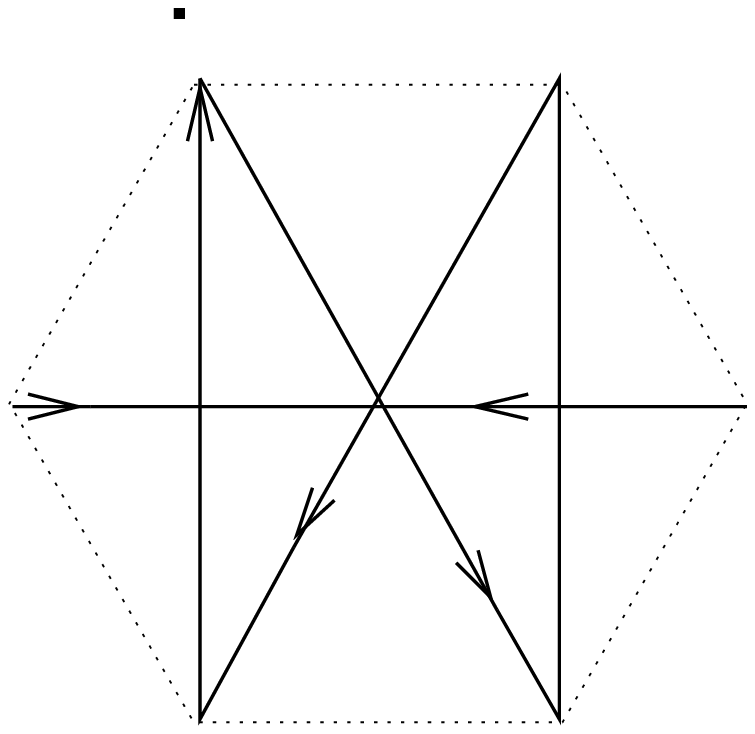}}
\end{picture}
\label{represent}
\caption{Diagrammatic way of representing the terms in eqs. 
(\ref{ex1}-\ref{ex2}).\label{exfiguur}}
\end{figure}

Let us consider some typical terms in the action at order~6:
\begin{eqnarray}
&&Tr(F_{\alpha_1 \alpha_2} F^{\alpha_2 \alpha_3} F_{\alpha_3 \alpha_4} F^{\alpha_4 
\alpha_1} F_{\beta_1 \beta_2} F^{\beta_2 \beta_1}) {\rm \ \ and} \label{ex1}
\\
&&Tr(F_{\alpha_1 \alpha_2} F_{\alpha_3 \alpha_4} F_{\beta_1 \beta_2} F^{\alpha_2 \alpha_3}
 F^{\alpha_4 \alpha_1} F^{\beta_2 \beta_1})\ . \label{ex2}
\end{eqnarray}
The interplay between the Lorentz index contractions and the group theory trace 
can be encoded in different ways. A pictorial way is to associate a diagram to each 
such term, by drawing points on the corners of a regular hexagon, indicating the 
position of the $F$-factors in the trace, and lines
(with arrows, which will however soon be dropped) connecting the different 
points, indicating the Lorentz contractions. The terms given in equations~(\ref{ex1}-\ref{ex2}) 
are then represented by figure \ref{exfiguur},
where the left most $F$ in the trace is represented by the upper left corner
of the hexagon.

An alternative description, geared towards the permutation group considerations
that follow, goes as follows. Label the first index on the 6 field strengths
from $1$ to $6$. Then the sequence of indices in the  second position
is a permutation of the first index. We will denote this permutation $i(.)$, and
use it to label the diagram. The permutations corresponding to
(\ref{ex1}-\ref{ex2}) are $(1234)(56)$ and $(1425)(36)$ in a cycle notation%
\footnote{i.e., for the second example, $i(5)=1,i(1)=4,i(4)=2$ etc.}.
Obviously, each term at order~6 can be represented by one (or more) of the
$6!=720$ possible permutations.

\subsection*{Conjugacy classes}
The complex linear combinations of diagrams are taken as a 
representation space for the permutation group of 6 elements.
The action of $S_6$ on this representation space is {\em by conjugation},
 as we now explain.
The action of the permutation group consists of 
reshuffling the vertices of the diagrams, which is the same as
reshuffling F's in the trace.  The action
of a permutation $g$ on the vertices becomes, after relabeling:
\begin{eqnarray}
g( F_{1, i(1)} \dots F_{6, i(6)} ) & \equiv &
 F_{g^{-1}(1), i (g^{-1}(1))} \dots F_{g^{-1}(6), i(g^{-1}(6))} \\
& = & F_{1, gig^{-1} (1)} \dots F_{6, gig^{-1} (6)}\,.
\end{eqnarray}
Evidently, the set of diagrams within one conjugation class is invariant
under this action. As far as this representation of the permutation
group on the diagrams is concerned, we can study each conjugation class 
separately. Each of these representations separately is in fact a (transitive)
representation by permutation of diagrams.

The arrows on the diagrams can be dropped.
Two diagrams that are the same up to the orientation
of a loop are equivalent \footnote{One can easily see that this
equivalence relation is compatible with the action of the group.}
since they correspond to the same term in the action (up to an unimportant
sign): reversing the arrow in a loop amounts to flipping the order of the indices 
in all the field strengths connected by that loop. 

\subsection*{An induced representation}
Now we analyse the representation of the permutation group on each conjugation class.
Consider a specific conjugation class, choose a diagram (without the arrows)
and label it $i_1$. 
The chosen diagram is invariant under a subgroup of the permutation group
(acting by conjugation as above). The invariance group 
of $i_1$ we call $H_1$. For both our examples (\ref{ex1}-\ref{ex2})
the invariance group is isomorphic with $\Z_4 \otimes \Z_2 \otimes \Z_2$.

It is clear that every other diagram $i$ in the conjugacy class can be reached 
by the action of some group element $g$, namely $i=gi_1g^{-1}$. 
Every $gh$ with $h \in H_1$ yields that same diagram $i$. Therefore the set 
of diagrams within a conjugacy class is the same as the set of
the left cosets with
respect to the invariance group (of a diagram in that conjugacy 
class). The action of the group on this set of cosets is the left regular 
action. This representation is the representation \em induced \em 
\cite{simon} by the 
trivial representation of $H_1$ on $S_6$. Via Fr\"obenius' character 
formula we can then decompose this induced representation in 
irreducible ones, using the character table of $S_6$. This decomposition
provides an inroad into the structure of 
the terms in the NBI, at order $F^6$ and potentially beyond.
Note that if we had picked a different diagam $i_2$ in the same conjugacy class
to start with, we would have $i_2=gi_1g^{-1}$ for some $g \in S_6$.
The invariance group  $H_2=gH_1g^{-1}  $ would yield an equivalent 
construction to the previous one. Therefore, $H_1 \cong H^{cc}$ is 
uniquely associated to a conjugacy class (c.c.).
The results for the invariance groups are summarized in 
table \ref{inv} \footnote{The symbol $\otimes$ in table \ref{inv}
does not denote a direct
product. It is easy to deduce from the context how the
product of subgroups should be taken. The subgroups are ordered as follows. First
the cyclic permutation within a loop, then the permutation of loops of equal length,
finally the orientation reversal.
For loops of length 2, this last group is trivial.},
for the relevant conjugacy classes\footnote{We momentarily explain
why other conjugacy classes are irrelevant.}.
The split into irreducible representations is
assembled in table \ref{split}.

\begin{table}[h]\begin{center}
\begin{tabular}{|c|c|} 
\hline 
Conjugacy class & Invariance group $H^{cc}$ \\
\hline
[ 6 ] & $ Z_6 \otimes Z_2 $ \\
\hline
[ 4 2 ] & $ Z_2 \otimes Z_4 \otimes  Z_2 $ \\
\hline
[ 3 3 ] & $ (Z_3)^2 \otimes S_2 \otimes (Z_2)^2  $ \\
\hline
[ 2 2 2 ] & $  (Z_2)^3 \otimes  S_3$ \\ \hline              
\end{tabular}
\caption{Invariance groups associated to conjugacy classes}    \label{inv}
\end{center}\end{table}

\subsection*{Cyclicity and double cosets}

At this stage a term in the Lagrangian corresponds to several diagrams,
since the trace is cyclic: a cyclic permutation corresponds to a rotation of 
the diagrams. We denote the subgroup of $S_6$ corresponding to these rotations 
as $N=Z_6^c$ (where $c$ stands for cyclicity).
Then it should be clear that the cosets $gH$ within the 
double coset $NgH$ correspond to equivalent diagrams. We finally obtain
therefore, that inequivalent diagrams correspond to double cosets $NgH$. 

To count these double cosets in the left regular representation on the 
$H$-cosets, it is sufficient to count $Z_6^c$ invariants within each 
irreducible component of the representation. To do that, we can use 
Fr\"obenius reciprocity and the character tables for $S_6$ and $Z_6^c$.

The results then at order~6 are the following.
Conjugacy classes of $S_6$ with a cycle of length 1, we do not consider 
since a field strength contracted with itself yields a  term equal to zero
in the action. We have only four 
conjugacy classes left then. The number of double cosets in each of these 
conjugacy classes is summarized in table \ref{split}, along with the decomposition
into irreducible representations.
\begin{table}[h]\begin{center}
\begin{tabular}{|c|c|c|}            
\hline
  Conjugacy class & Irreducible reps & \# invariants \\
\hline [6]  &  $ 60= [6]+[21^4]+2.[2^3]+2.[42]+[31^3]+[321] $ & 14  \\
\hline [42] &  $ 45= [6]+[51]+[2^3]+2.[42]+[321]              $  & 9   \\ 
\hline [33] &  $ 10= [6]+[42]            $ & 3   \\ 
\hline [222]  & $ 15= [6]+[2^3]+[42]             $  & 5   \\
\hline                          
\end{tabular}
\caption{Irreducible components and double cosets}      \label{split}
\end{center}\end{table} 
The diagrams corresponding to these invariants are drawn and labelled in 
appendix \ref{diagrams}.

As we already indicated, 
this analysis generalizes to any order and gives therefore 
a systematic way to count the number of unknown coefficients in the NBI action
including terms written using  field strengths only, at any given order.

\subsection*{Invariant linear combinations}

In the previous analysis, we split the representation space into conjugacy
classes, next into
inequivalent irreducible representations, and we determined the number of 
double cosets within an irreducible representation. Now we would like to 
write down explicitly these $Z_6^c$ invariants in terms of the diagrams, 
which translates directly into terms in the action at order~6. 

For most of the irreducible representations, the number of corresponding 
invariants is larger than one. Lacking a criterium to decide which linear 
combinations are most suitable, we made the following arbitrary choice.
Corresponding to a specific irreducible representation of the permutation 
group, there is a Young diagram and a Young symmetrizer: acting with this Young 
symmetrizer on a specific diagram yields automatically a vector in that irrep.
The resulting vector can then simply be symmetrised with respect to the $Z_6^c$ 
cyclic group. The result of this procedure is one of the sought after 
invariants. We have recorded in table~\ref{invariants} (in the appendix~\ref{BigTable})
a complete%
\footnote{We have not bothered to include the results of this analysis for the 
terms with Lorentz contraction structure [3 3]. The reason is that, for the 
backgrounds we have studied, these terms give {\em no} contribution to the 
quadratic action for the fluctuation, and therefore these terms remain 
completely arbitrary.\label{voetnoot33}}
 set of combinations obtained in this way, together with the $S_6$ 
irrep in which they are found. Each line involves a choice of starting diagram,
for which we found no good criterium
(like an a priori guarantee to give a linearly independent combination).

Alternatively, one may project on a generically reducible subspace formed by the
sum of equivalent $S_6$ representations using the minimal projection 
operator \cite{simon} $e({\cal F})$ associated to a specific irrep ${\cal F}$.
For example in the $[6]$ class the projection
on the $[42]$ representations yields a reducible representation, $[42]\oplus[42]$.
Each of these contains two $Z_6^c$ invariants. 
Acting with $e({\cal F})$ on a few (arbitrarily chosen) diagrams yields vectors
from this reducible space, and it is easy to pick out specific $Z_6^c$ invariants. 
Since this seems to offer no particular advantages (each choice of basis
for the resulting invariants seems arbitrary), we do not dwell on this further.

We pause here a moment to return to the results obtained in section 
\ref{order4}. If we carry out the group theory analysis described
in the previous paragraphs at order 4,
we obtain the  results in table \ref{sum4}.

\begin{table}[h]\begin{center}
\begin{tabular}{|c|c|c|c|c|}
\hline
Conjugacy class & irrep  & Linear 
combination & Name \\ \hline
[4] & [4]   & $ i_1+2i_2 $ & $ I_1^{4} $ \\ \hline
[4] & [2 2]  & $  i_1-i_2 $ & $ I_2^{4} $ \\ \hline
[22] & [4]  & $ i_3+2i_4 $ & $ I_1^{22} $ \\ \hline
[22] & [2 2]&$  i_3-i_4$ & $ I_2^{22} $ \\ \hline
\end{tabular}
\caption{Combinations of diagrams based on the permutation group: order~4. 
The square diagram is represented by $i_1$, the diabolo is $i_2$, the cross is  $i_3$ and 
$i_4$ the two parallel lines.
} \label{sum4}
\end{center}\end{table}
In the part of the NBI action purely in terms of field strengths at order four
only two of the four potential fourth--order invariants are actually present:
\begin{eqnarray} 
{\cal L}_{NBI-F^4} &=& \frac{1}{4} I_2^2  + \frac{1}{24}( I_1^4 - \frac{1}{4} I_1^{22}).
\end{eqnarray}
The group theory that we introduced will similarly simplify the form of 
the NBI action at order~6. At this stage we performed only the first step,
providing a catalogue of combinations that 
are in the different irreps, as recorded in table~\ref{invariants} in the appendix.
We now proceed to impose the data from the known string spectra.

\section{A NBI at order 6}
\label{order6}

\paragraph{Reality}                                  \label{reality}
A first, fairly trivial constraint on the action comes from
the demand that the action be real. The complex conjugate of a term
represented by a diagram, is given by the term corresponding to the mirror
diagram. This can easily be seen using the hermiticity of the Lie algebra
generators. We conclude that
diagrams that are mirror to each other have complex conjugate 
coefficients. The diagrams that are mirrorsymmetric have  real 
coefficients \footnote{This mirror-operation is the only group operation 
represented on all double cosets.  }.

Note that all diagrams at order 4 were 
mirrorsymmetric, and therefore they all necessarily had real coefficients. 
This is not true at sixth order. However, it turns out that, apart from the 
general structure as described for the fourth order calculation in 
equation~(\ref{structure2beYM}), an additional term is present at sixth order, 
that is off-diagonal in the $SU(2)$ components  of the field fluctuations.
The rescaled Yang--Mills requirement  puts this to zero.
This annihilates the imaginary parts of the complex conjugate coefficients
so that, as a conclusion, also at sixth order {\em all} coefficients are real.

\paragraph{String spectrum data}
In section \ref{order4} we executed our program of demanding a rescaled 
Yang-Mills action for the action quadratic in the fluctuations on our 
testing ground. It was succesful there in determining the coefficients
of the NBI at order 4 that we know to be correct. In this section, as discussed 
previously, we explore which constraints are found on the NBI if 
we extend this analysis to order~6.
 
The action for the quadratic fluctuations  at order~6 has virtually the same
structure as that discussed in detail for order~4 in section \ref{order4}. 
We follow the same route and rescale the action by $c_i^{\rm kin}$
and demand that the action is a rescaled YM action with appropriate rescaling factor.
The constraints from gauge invariance (see section~\ref{order4})
were not imposed a priori, but were used as a check on the computation.
The result is a large set of linear equations for the coefficients $a^{nnn}_{m}$ of the different 
terms in the action (see eq.~\ref{genact}). Of these, 21~are independent, 
leaving 10~out of~31 (see table~\ref{split}) of the coefficients in the general sixth order action undetermined.

Of these~10 undetermined coefficients,
3~are the coefficients of the invariants in class [3 3]: for the 
background we consider these invariants give vanishing contribution as we now argue%
\footnote{ It is 
 obvious that this argument extends to very many higher order terms with a structure that 
 factorises with Lorentz contractions of an odd number of field strengths.}.
The background (see section~\ref{plan}) has block-diagonal 
fieldstrengths, and therefore the Lorentz contraction of three background fields 
is frustrated and vanishes%
\footnote{The same argument eliminates terms arising from $\delta_2F$, see 
equation~\ref{delta1}.}.
Consequently, the quadratic variations could only 
arise when the Lorentz contraction structure is
$({\cal F F}\delta_1F)$ times $({\cal F F} \delta_1F)$.
But this also vanishes, since the $k-$ sum in
${\cal F}^i_k  {\cal F}^k_j$ will contain only one term, and is hence diagonal in $ij$. 
We ignore this terms in the sequel, and continue with the remaining 28~terms of 
table~\ref{invariants} in appendix~\ref{BigTable},
7~combinations out of~28 having arbitrary coefficients.

To present the result in detail, we make a change of basis. We still base our 
choice of combinations of diagrams on the permutation group considerations of 
section~\ref{group}. We remind the reader that in many 
cases, a given irrep occurs more than once, and in addition a given irrep usually 
contains two invariants. For such cases, the choice of basis for specific 
invariants is a priori quite arbitrary, and what was written in 
table~\ref{invariants} is a `raw' choice. With hindsight, this choice can be 
improved, and the result is recorded in table~\ref{BigTableToo}. The following 
changes were made:
\begin{itemize}
\item If the value of coefficients for a given representation is completely 
fixed%
\footnote{We are here taking into account the requirements from reality and
the rescaled Yang--Mills ansatz, not the `BPS' conditions. See further for the
 incorporation of those.}%
, this combination was chosen as one of the basis vectors. The other basis 
vectors (which therefore have zero coefficients) were taken to be orthogonal 
with the natural metric for the diagrams. This is the case for
the $[4 2]$ and $[321]$ irreps in all classes, as well as for the $[6]$.
Whereas this last fact is obvious (it corresponds to the abelian case), the 
general reason for the other ones is unclear.
\item If the values of the coefficients are fixed numbers for some, and 
arbitrary parameters for other combinations, we have separated the basis accordingly.
This is the case for all the $[2^3]$ irreps, where for each class a single 
combination is fixed, and for the $[3 1^3]$ likewise. 
\item The stand--alone $[2 1^4]$ invariant (which has arbitrary coefficient as well)
is not touched.
\end{itemize}

\begin{center}
\begin{table}
\begin{tabular}{|c|c|c|c|c|}
\hline
Class & $S_6$-rep & Invariant linear combination & coefficient &name \\
\hline
 2 2 2 & $ [6]   $ & $  3 i_1 + 2 i_2 + 1 i_3 + 3 i_4 + 6 i_5 $ & abel &$I_6^{222}$ \\  \hline
 2 2 2 & $ [4 2] $ & $  1 i_1 + 4 i_2 - 3 i_3 - 4 i_4 + 2 i_5 $ & fixed&$I_{42}^{222} $ \\ \hline
 2 2 2 & $ [4 2] $ & $  2 i_1 + 1 i_2 + 1 i_3 - 1 i_4 - 3 i_5 $ & 0  & \\  \hline  
 2 2 2 & $ [2^3] $ & $ -3 i_1 + 2 i_2 + 1 i_3 + 0 i_4 + 0 i_5 $ & fixed &$I_{222}^{222}$ \\  \hline
 2 2 2 & $ [2^3] $ & $  0 i_1 - 1 i_2 + 1 i_3 - 3 i_4 + 3 i_5 $ & undet&$I_{222}^{'\,222}$ \\ \hline \hline 
 4 2 & $ [6] $ & $ 2,2,2,2,2,2,1,1,1$  &abel& $I_6^{42} $ \\  \hline  
 4 2 & $[4 2]$ & $2,\,0\,,2,1,-1,-1,\,0\,,-2,-1 $&fixed&$I_{42}^{42}$\\ \hline  
 4 2 & $[4 2]$ & $-1,3,2,-2,-1,-1,\,0\,,1,-1$ &0&\\ \hline    
 4 2 & $[4 2]$ & $3,3,-2,-2,3,3,-6,-1,-1$ &0&\\ \hline                                                      
 4 2 & $[4 2]$ & $-6,\,0\,,1,4,3,3,\,0\,,-1,-4$ &0&\\ \hline
 4 2 & $[3 2 1]$ & $-2,\,0\,,2,-2,1,1,\,0\,,-2,2$ &fixed&$I_{321}^{42}$\\ \hline
 4 2 & $[3 2 1]$ & $\,0\,,\,0\,,\,0\,,\,0\,,1,-1,\,0\,,\,0\,,\,0\,$ &0&\\ \hline 
 4 2 & $[2^3]$ & $ \,0\,,2,-2,\,0\,,\,0\,,\,0\,,1,-1,\,0\,$ &fixed&$I_{222}^{42}$\\ \hline
 4 2 & $[2^3]$ & $1,-1,\,0\,,-2,1,1,1,\,0\,,-1$ &undet&$I_{222}^{'\,42}$\\ \hline\hline
 6 & $[6] $ & $ 1,6,6,3,6,6,6,6,2,3,6,3,3,3$ &abel& $I_6^{6} $ \\ \hline  
 6 & $[4 2]$ & $1,2,-2,-1,-2,-2,-6,-2,-2,3,2,5,3,1$&fixed&$I_{42}^6$\\ \hline  
 6 & $[4 2]$ & $1, 2, 1, 2, 1, 1, -3, -2, 1, 0, -1, -1, 0, -2$ &0&\\ \hline    
 6 & $[4 2]$ & $1, 6, 1, -2, 1, 1, 1, 6, -3, -2, -9, 3, -2, -2$ &0&\\ \hline                                                      
 6 & $[4 2]$ & $2, -3, -4, 5, -4, -4, 2, 3, 3, -1, -3, 3, -1, 2$ &0&\\ \hline
 6 & $[3 2 1]$ & $2,2,1,0,1,-4,2,-2,-2,-1,2,-2,-1,2$ &fixed&$I_{321}^6$\\ \hline
 6 & $[3 2 1]$ & $0,0,-1,0,1,0,0,0,0,1,0,0,-1,0$ &0&\\ \hline 
 6 & $[2^3]$ & $1, 2, -2, -3, -2, 2, 2, -2, 2, -1, 2, 1, -1, -1$ &fixed&$I_{222}^6$\\ \hline
 6 & $[2^3]$ & $1, -4, 1, 0, 1, 5, -1, -2, -1, -1, -1, 1, -1, 2$&undet&$I_{222}^{'\,6}$\\ \hline
 6 & $[2^3]$ & $0, 2, -2, 0, -2, 2, 2, -2, 0, 2, -4, -2, 2, 2$ &undet&$I_{222}^{''\,6}$\\ \hline                                                     
 6 & $[2^3]$ & $0, -2, 2, 0, 2, -2, 4, -4, 0, 1, -2, 2, 1, -2$ &undet&$I_{222}^{'''\,6}$\\ \hline
 6 & $[3 1^3]$ & $0,0,-2,0,2,0,0,0,0,-1,0,0,1,0$ & fixed=0 &$I_{3111}^6$\\ \hline 
 6 & $[3 1^3]$ & $1,-2,-1,0,-1,1,1,2,-1,1,1,-1,1,-2$ &undet &$I_{3111}^{'\,6}$\\ \hline                                                      
 6 & $[2 1^4]$ & $ 1,-2,2,-3,2,-2,-2,2,2,1,-2,-1,1,1$ & undet &$I_{21111}^{'\,6}$\\ \hline
\end{tabular}
\caption{Results on the coefficients of cyclic invariants by irreducible representation.
The last column contains names for future reference.
The column before last has the following meaning: `abel' indicates a coefficient fixed
by the abelian case (or by Tseytlin's proof of the `symmetric trace' formula),
`fixed' and `undet' mean fixed resp. undetermined by the rescaled Yang--Mills 
analysis. } 
\label{BigTableToo}
\end{table}
\end{center}
In table~\ref{BigTableToo}, we have again listed the potential cyclic 
invariants in the  group-theoretic classification, in the changed basis.
The resulting sixth order terms in the action are
\begin{eqnarray}
{\cal L}^{(6)}&=&
\frac{1}{720} I_6^{6}+ \frac{1}{6480} I_{42}^6 -\frac{1}{5760} I_{321}^6 + \frac{1}{720} I_{222}^6
\nonumber\\
&&
+ \lambda_1\,  I_{222}^{'\,6} + \lambda_2\,  I_{222}^{''\,6}+ \lambda_3\,  I_{222}^{'''\,6}
 + \lambda_4\,  I_{3111}^{'\,6} + \lambda_5\,  I_{21111}^{'\,6}
\nonumber\\
&&
 -\frac{1}{480} I_6^{42} + \frac{1}{3240} I_{42}^{42}  -\frac{1}{11520} I_{321}^{42}  
  -\frac{1}{360} I_{222}^{42}
   + \lambda_6\,  I_{222}^{'\,42}
\nonumber\\
&&
 +\frac{1}{5760} I_6^{222} -\frac{1}{25920} I_{42}^{222} -\frac{1}{2880} I_{222}^{222}
 +  \lambda_7\,  I_{222}^{'\,222}\,.
\label{ActionSixthOrder}
\end{eqnarray}

It is clear that the resulting expression displays a remarkable amount of 
structure, but we have not been able to penetrate beyond the obvious.

An important check on the arbitrariness is provided by the fact that some 
commutator combinations can not possibly contribute in the restricted class of 
background that we investigated. These are, in an obvious notation (see 
appendix~\ref{notationsappendix} if an explanation is needed)
\begin{eqnarray}
&& Tr [F_1 F_{1'}][F_2 F_{2'}] [F_3 F_{3'}]\,, \\
&& Tr [F_1 F_{1'}][F_2 F_{2'}] [F_3 F_4]\,, \\
&& Tr [F_1 F_{2}][F_3 F_{4}] [F_5 F_6]\,, \\
&& Tr [F_1 F_{2}][F_3 F_{5}] [F_4 F_6]\,, \\
&& Tr [F_1 F_{2}][F_2 F_{5}] [F_4 F_6]\,. 
\end{eqnarray}
The reason is obvious: the quadratic 
variation of these products of three commutators always has one commutator left,
and vanishes since the background is abelian.%
\footnote{It is less obvious that the (independent) combination $F_{1}F_{[2}F_{|3|}F_4F_5F_{6]}$
(leaving the label 3 out of the antisymmetrisation) also does not contribute: 
here the block-diagonal nature of the background is involved.
This is in fact the invariant $I_{21111}^{'\,6}$ with coefficient $\lambda_5$.
We have no short explanation for the seventh invariant, with $\lambda_3$.}
The first line corresponds to~$\lambda_7$,
the second to~$\lambda_6$. The last three lines generate through 
linear combinations the $[3111]$--invariant with $\lambda_4$, as well as the $[222]$ 
invariants with $\lambda_1$ and $\lambda_2$.

\section{BPS configurations}
\label{BPS}
Turning on magnetic fields usually breaks all supersymmetry with as a 
result that the D-brane configuration becomes unstable. 
This can already be seen from the massformulae, 
eqs. (\ref{stringspec}) and (\ref{ymspec}), which exhibit
the generic presence of tachyonic modes in the spectrum. However, it was 
noticed in \cite{BDL} that for very specific choices of the background 
some supersymmetry survives. 

We will first formulate this in the T-dual picture. We take two D$p$-branes,
one of them in the $(2,4,\cdots, 2p)$ direction and the other one rotated 
over an angle $\gamma_1$ in the (2\,3) plane, over an angle $\gamma_2$ in the 
(4\,5) plane, ..., over an angle $\gamma_p$ in the $(2p\,2p+1)$ plane. Searching 
for common directions in the supersymmetry charge and the rotated 
charge gives BPS configurations which are summarized below.

\begin{center}
\begin{tabular}{|c|l|c|l|}\hline\hline
$p$ &BPS angle & susy's& BPS magnetic fields\\ \hline\hline
2 &$\gamma_1=\gamma_2$ & 8&$f_1^3=f_2^3$\\ \hline
3 &$\gamma_1=\gamma_2+\gamma_3$ & 4&
$f_1^3=f^3_2+f^3_3+f^3_1f^3_2f^3_3$\\ \hline
4 &$\gamma_1=\gamma_2+\gamma_3+\gamma_4$ & 2&
$f^3_1=f^3_2+f^3_3+f^3_4+f^3_1f^3_2f^3_3+$\\
&&&$\quad f^3_1f^3_3f^3_4+f^3_1f^3_2f^3_4-f^3_2f^3_3f^3_4$
\\ \cline{2-4}
  &$\gamma_1=\gamma_2$, $\gamma_3=\gamma_4$&4 &
$f^3_1=f^3_2$, $f^3_3=f^3_4$  
\\ \cline{2-4}
  &$\gamma_1=\gamma_2=\gamma_3=\gamma_4$&6&
$f^3_1=f^3_2=f^3_3=f^3_4$  
\\ \hline\hline
\end{tabular}
\end{center}

\noindent We assumed that none of the angles are zero. 
In the table we list the conditions on 
the angles, the number of preserved supercharges and finally the 
T-dual picture where the condition on the angles is translated, using 
eqs. (\ref{relang}) and (\ref{relang0}), 
into a condition on the magnetic fields. For simplicity we took the 
magnetic field entirely in the $\sigma_3$ direction, i.e. $f_i^0$ in equation~(\ref{8}) vanishes.

Though the conditions on the angles are linear, they translate for two cases 
into non-linear conditions on the magnetic fields. 
In fact, when switching on the $U(1)$ part of the magnetic 
field, one always gets such corrections.  At first sight one would expect 
this to give a crucial handle on the NBI. Indeed, BPS configurations 
should solve the equations of motion with as a result that the non-linear 
conditions relate different orders in the NBI. However all backgrounds 
considered above are in the torus of $U(2)$ and thus insensitive to 
different ordenings in the equations of motion. In fact they all solve the equations 
of motion of the abelian Born-Infeld action and as a consequence those arising from
our action through order $F^6$ as well. There is one case where we do 
have a good guess for the general BPS condition: rotated D2-branes or 
D4-branes with magnetic fields. In that case the obvious guess for
the full non-abelian BPS condition is self-duality of the magnetic field.

In \cite{brech}, some arguments were put forward to sustain the
claim that self-dual static magnetic 
backgrounds solving the equations of motion while 
simultanously minimizing the energy is equivalent to demanding that the whole NBI for such 
configurations collapses to the leading Yang-Mills term. It was shown that the symmetrized
trace prescription does share this property. Implementing this assumption 
in our case gives five conditions on the general form of the action at sixth order.
It turns out that three of these are dependent on the previously implemented rescaled
Yang--Mills conditions, thus providing a consistency check on both our 
results and the proposal in \cite{brech}.
The remaining two take an extremely simple form 
in terms of the coefficients in eq.~(\ref{ActionSixthOrder}), viz.
\begin{eqnarray}
\lambda_3&=&\frac{1}{1440}\nonumber\,,\\
\lambda_7&=&\lambda_1-\frac{\lambda_2}{4}+\frac{\lambda_6}{2}\,.\nonumber
\end{eqnarray}
Note that different Lorentz contraction structures are connected\footnote{
Selfduality, $F_{\mu \nu}=\varepsilon_{\mu \nu}{}^{\rho\sigma}F_{\rho\sigma}/2$,
implies that $F_{\mu \rho} F^\rho{}_\nu+ F_{\nu \rho} 
F^\rho{}_\mu=\eta_{\mu \nu} F_{\rho\sigma}F^{\sigma\rho}/2$. Repeatedly using this, 
allows us to rewrite all terms of class 6 and 4 2 in terms of the five elements 
of the class 2 2 2.}.
As far as the permutation group structure is concerned, the conditions are pure 
$[2 2 2]$.

What about the full non-abelian version of the non-linear BPS conditions? While the 
first order correction to the linear relations can easily be deduced from 
the fact that they should solve the equations of motion through order 
$F^3$, nothing can be said to all orders yet. 
A more detailed study of these BPS configurations and their consequences for the NBI, 
has to wait for a better 
understanding of supersymmetry in the NBI \cite{BDSip}.

\section{Conclusions}
In this paper, we made a first systematic
attempt to determine corrections to the 
Str-terms in the NBI action. The physical testing ground on which we 
worked, were D-branes wrapped on tori with magnetic backgrounds turned on. 
Central in an important part of our analysis is the fact that the spectrum for open 
strings stretching between these D-branes as predicted by perturbative 
string theory differs by a mere rescaling from the YM-approximation to the 
spectrum. We made two bold assumptions to proceed in unknown territory. 
The first was that we did not take along \em all \em possible derivative 
corrections to the NBI. This was inspired partly by practical motives, 
partly by the second assumption, namely, that the action quadratic in the 
fluctuations in this background obtained from the NBI should be a rescaled
YM action. Under these assumptions we were able to put severe 
constraints on the NBI action. A weak a posteriori argument is that this 
method yields correct results at order 4. More encouraging is the fact that 
this approach yields  constraints that are compatible
with the constraints we obtained from analysing BPS configurations
-- this was not evident from the outset.

By construction, this action heals a severe default of the Str NBI action, 
pointed out in \cite{HT} \cite{B} and \cite{DST}, namely that it doesn't 
predict the correct spectrum for open strings on our testing ground. A
direct calculation of the terms at order 6 of the NBI via six point 
functions in string theory or a five loop beta-function calculation in a 
non-linear $\sigma$--model would be welcome, of course, to see whether our 
assumptions are valid. As long as this calculation is not available, other methods to 
get a grip on the NBI are worth study. A natural extension of our ideas is to 
just enlarge the testing ground by looking at other compactification 
manifolds, and by looking at different, possibly electric backgrounds. It 
could help also if we could gain more insight into the relative norms of 
the linear combinations of diagrams corresponding to double cosets, for instance,
to see 
whether there might be a systematic expansion for the NBI -- although this 
is perhaps asking for too much \footnote{We thank J. de Boer for a 
discussion on this point.}.

There are of course alternative techniques that are more complementary to 
our approach. The most promising route to obtain a grip on the 
non-abelian Born-Infeld might be via supersymmetry. Simply by noethering 
one can try to modify ten dimensional YM with non-linear 
corrections and modify the supersymmetry transformation rules accordingly. 
In the abelian case this fixes e.g. uniquely the fourth order term in the BI action 
\cite{BMT}. Because of the severely restricted form of the supersymmetry 
algebra in ten dimensions, it might even be%
\footnote{This was suggested to us by Savdeep Sethi.}
that the BI action is the only 
supersymmetric deformation of abelian YM. Continuing this line of thought, 
it should be clear that a similar analysis should be performed in the 
non-abelian case. A good starting point would be the BPS conditions in higher 
dimensions in section \ref{BPS} -- they provide a hint of how to modify the
supersymmetry variations. It seems important to us to carry out 
this program in the maximum of ten dimensions, because the supersymmetry 
algebra is largest there and therefore puts (much) stronger restrictions
on the form of the action. For lower dimensions some partial
results for a supersymmetric non-abelian extension of Born-Infeld theory
are available \cite{zanon}.

Closely linked to the idea of a supersymmetric action on the brane is the 
idea of the construction of an action invariant under $\kappa$--symmetry \cite{BdS}. 
This approach starts from the observation that the form of the Wess-Zumino term, 
which describes the coupling of the gauge fields to the Ramond-Ramons bulkfields,
is severely restricted, even in the non-abelian case. As the variation of 
the Wess-Zumino term under the $\kappa$-transformations has to be 
cancelled by the variation of the NBI, one gets a recursive method to 
construct the NBI. This program was already carried out through quartic order
in the Yang-Mills field strength, and including all fermion
bilinear terms up to terms cubic in the field strength \cite{BdS}. The ordenings are 
indeed completely fixed by requiring $\kappa$-invariance. Surprisingly, 
it was found that at such low order, deviations of the symmetrized trace 
proposal do already appear. 

Another route to the NBI derives 
from the study of the equivalence of non-commutative and
commutative Born-Infeld actions via the Seiberg-Witten map. In this way one
obtains constraints
on derivative corrections to the Born-Infeld action. It would
be interesting to see whether this can teach us anything 
about the NBI action, as claimed in \cite{CS}.

Solutions to different 
non-abelian extensions of Born-Infeld theory, not necessarily related to string
theory have been studied.
The non-linearity of the action often leads to a 
smoothing out of solutions to an ordinary YM or Maxwell action. 
It could be interesting to study what kind of corrections can be expected 
from more general proposals for non-abelian extensions of Born-Infeld 
theory, including ours.

\section*{Acknowledgements}

We thank Eric Bergshoeff, Jan de Boer, Mees de Roo, Savdeep Sethi
 and Pierre van Baal
for useful discussions. A.S. and W.T. are supported by the European Commission
RTN programme HPRN-CT-2000-00131, in which A.S. is associated
to the university of Leuven.
The work of J.T. is supported in part by funds provided by the U.S. Department
of Energy under cooperative research agreement DE-FC02-94ER40818. J.T.
moreover thanks the Vrije Universiteit Brussel and the FWO Vlaanderen
 for support
during the first stages of this work.

\appendix

\section{Drawing diagrams}
\label{diagrams}
We draw here all the diagrams corresponding to the 
double cosets as introduced in section \ref{group}. \footnote{In fact,
it was intuitively easy to see, even before we knew the group theory from section 
\ref{group} that drawing all different diagrams was sufficient to enumerate 
all different terms in the action.}
\begin{figure}[h]
\begin{minipage}[t]{\textwidth}
\hfill\hfill
\includegraphics[angle=-90,scale=.2]{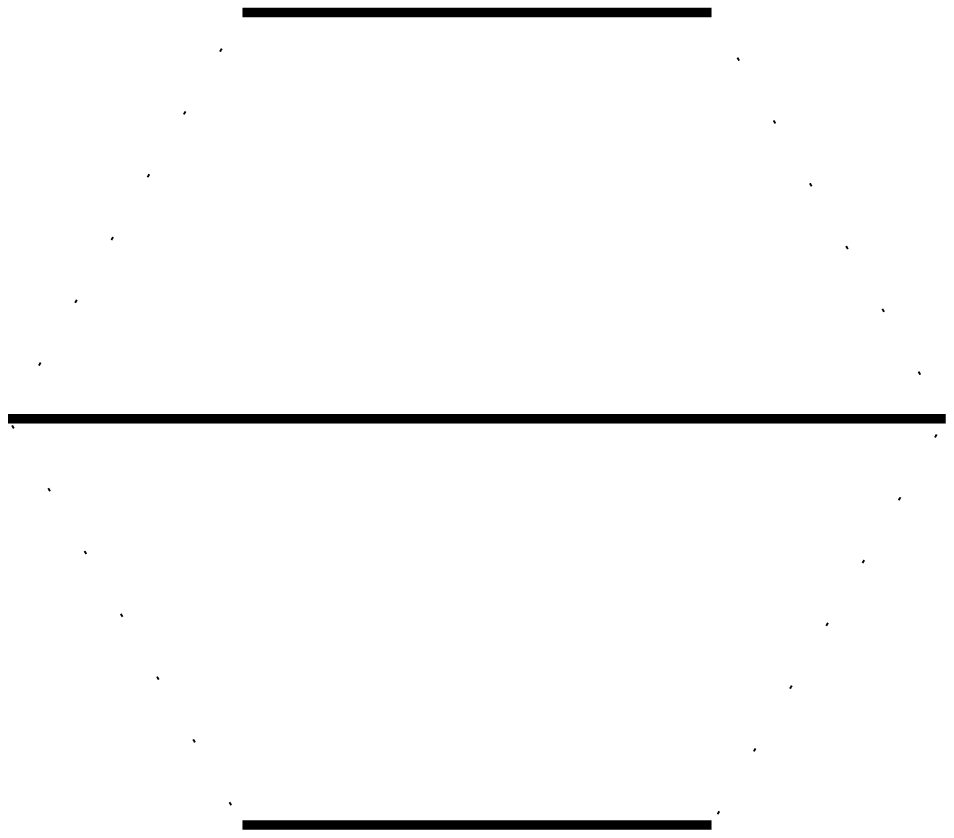}
\hfill
\includegraphics[angle=-90,scale=.2]{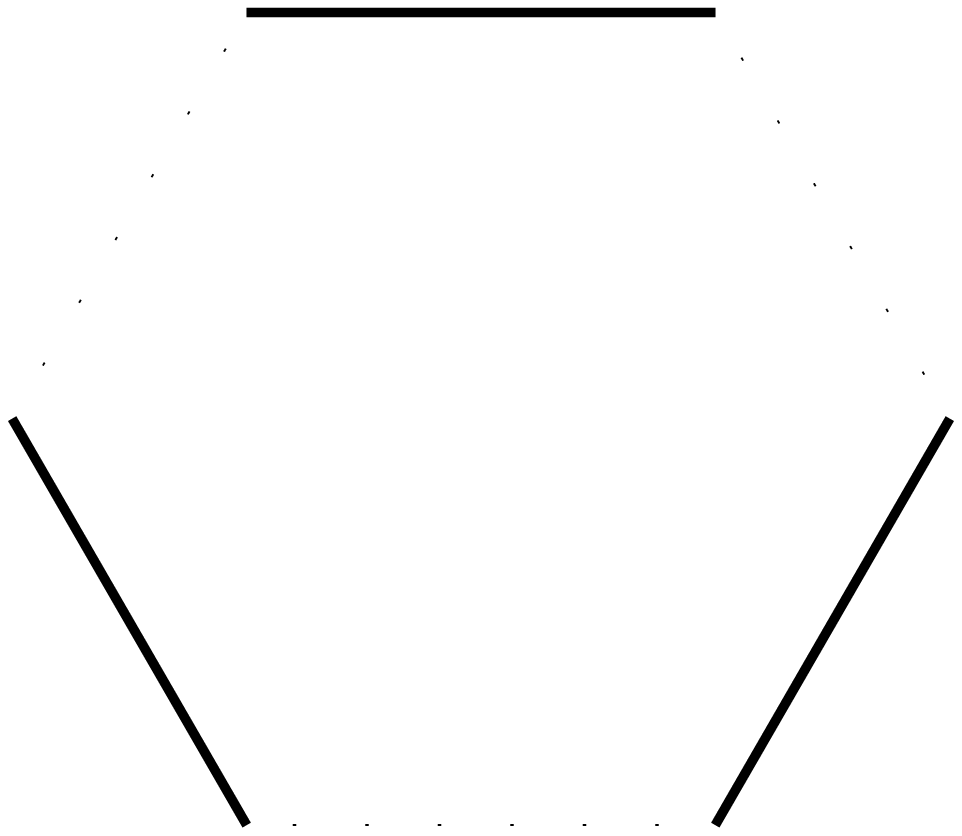}
\hfill
\includegraphics[angle=-90,scale=.2]{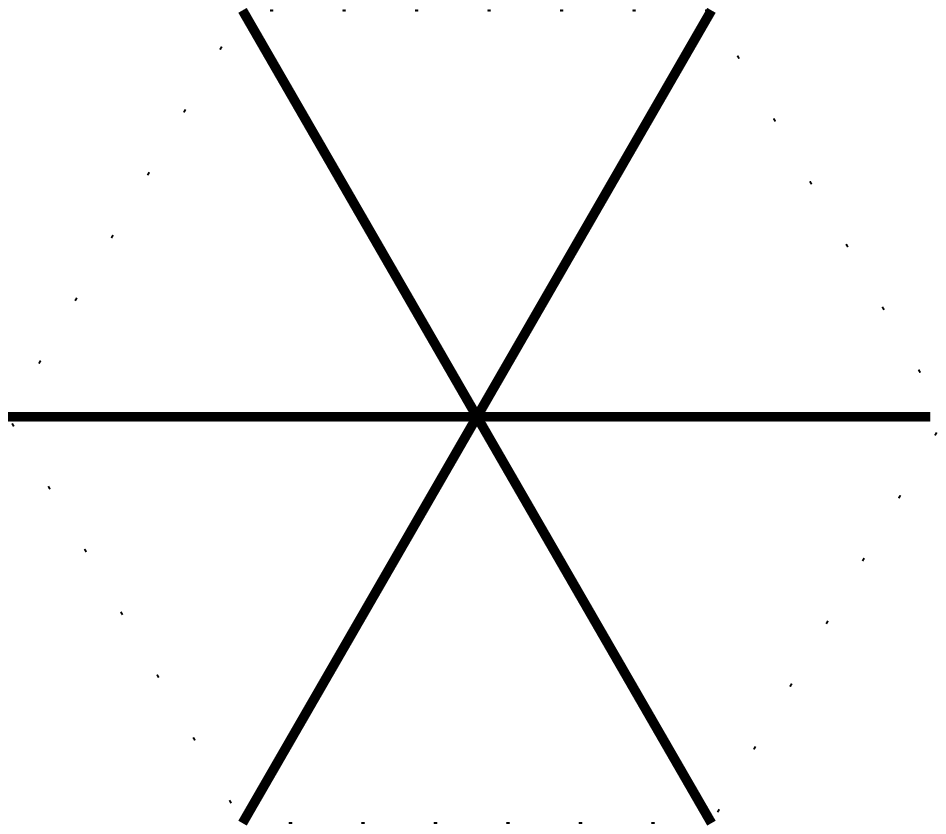}   
\hfill
\includegraphics[angle=-90,scale=.2]{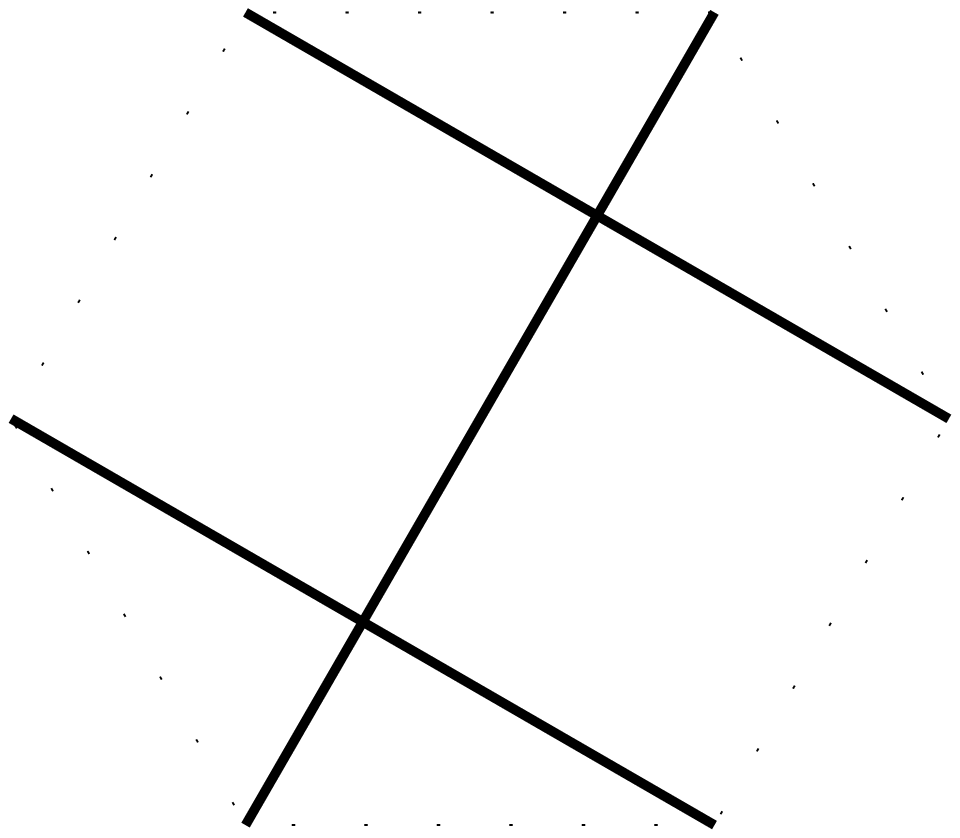}
\hfill
\includegraphics[angle=-90,scale=.2]{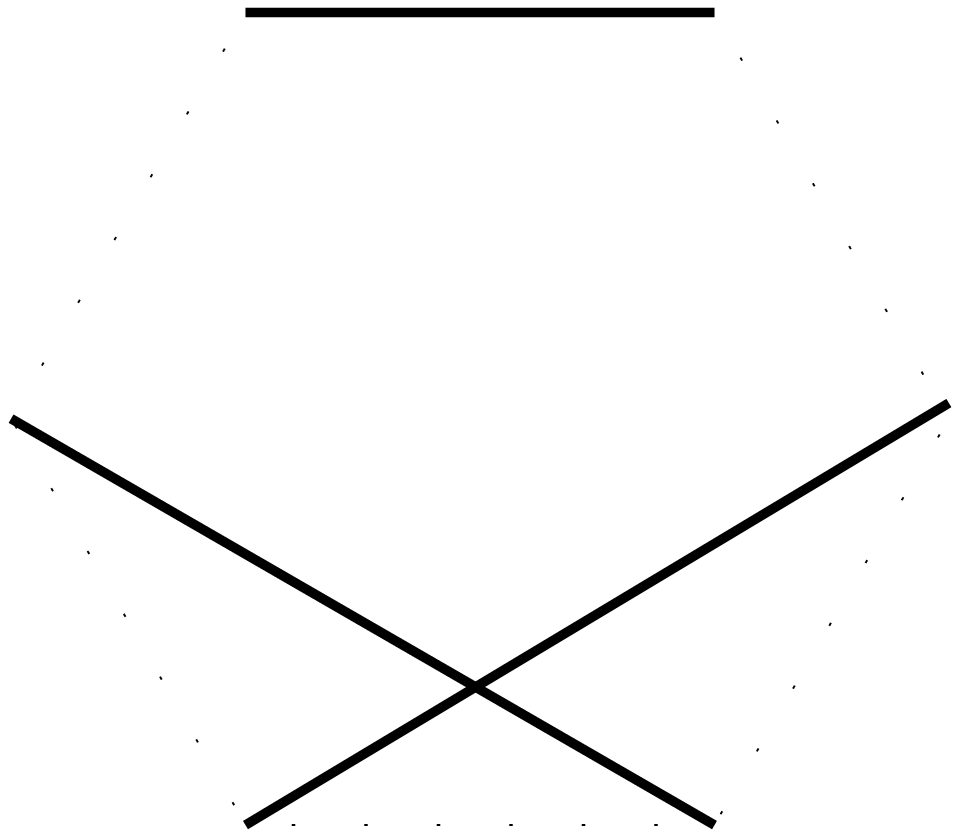}   
\hfill
\end{minipage}
\caption{Diagrammatic  representation of the five [222] terms in the action.} 
\label{vijf keer 222}
\end{figure}
\newline
\begin{figure}[h]
\begin{minipage}[t]{\textwidth}
\hfill \raisebox{-.07\textwidth}{1}
\hfill
\includegraphics[angle=-90,scale=.2]{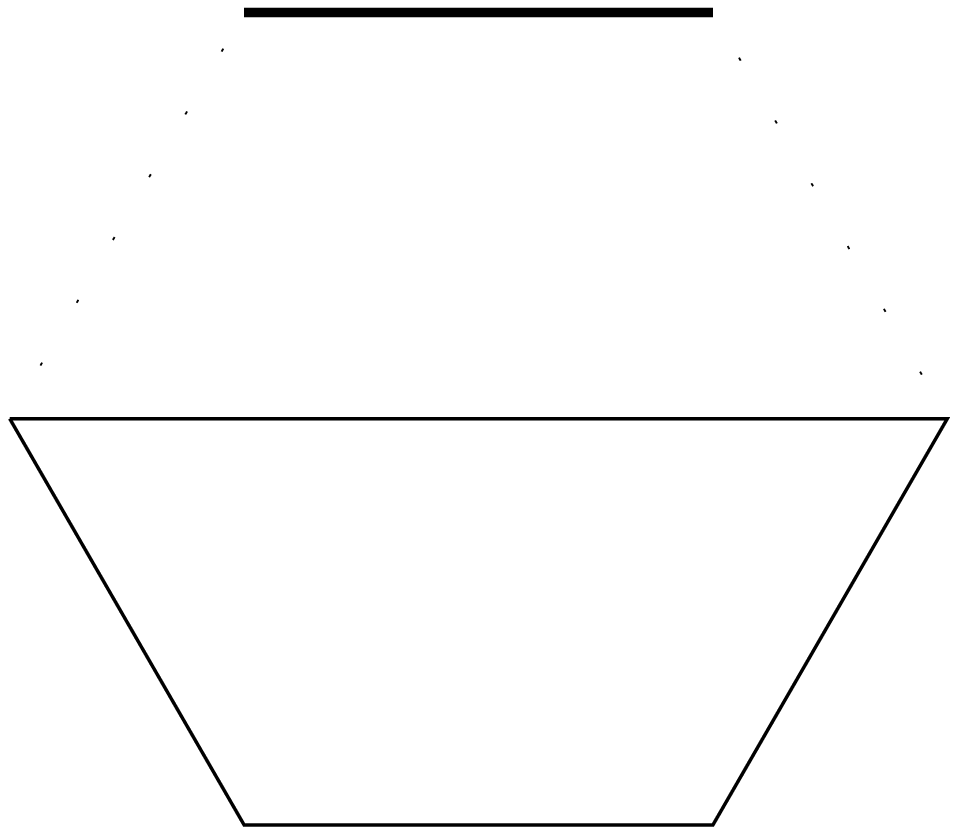}
\hfill
\includegraphics[angle=-90,scale=.2]{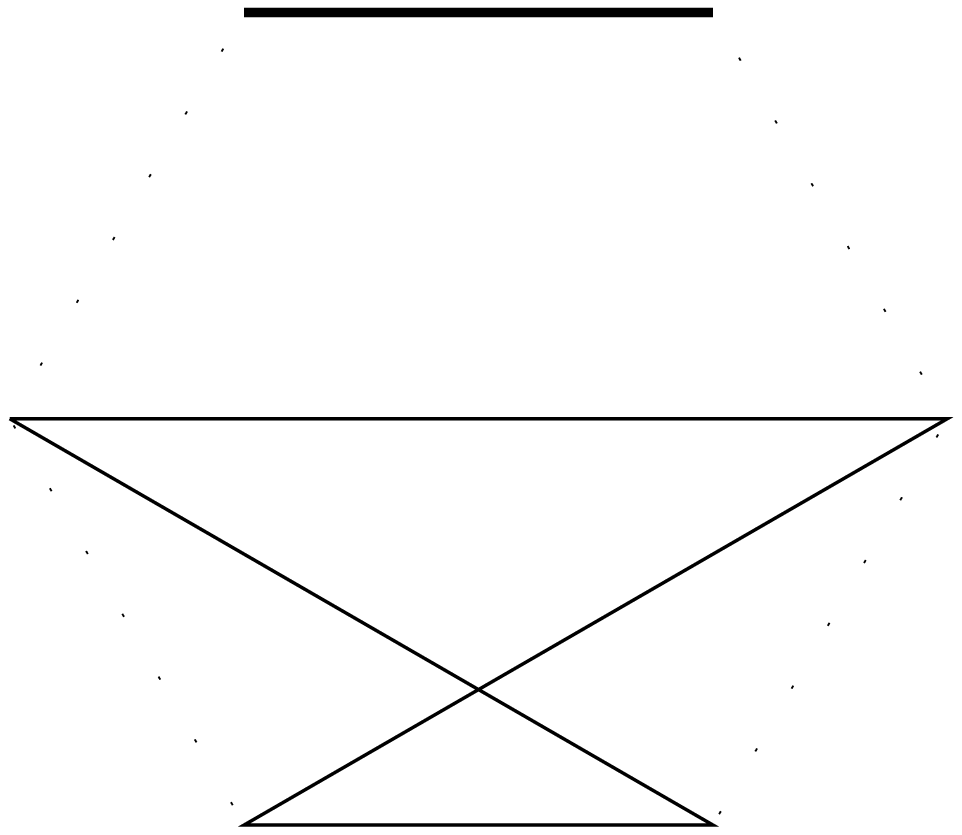}
\hfill
\includegraphics[angle=-90,scale=.2]{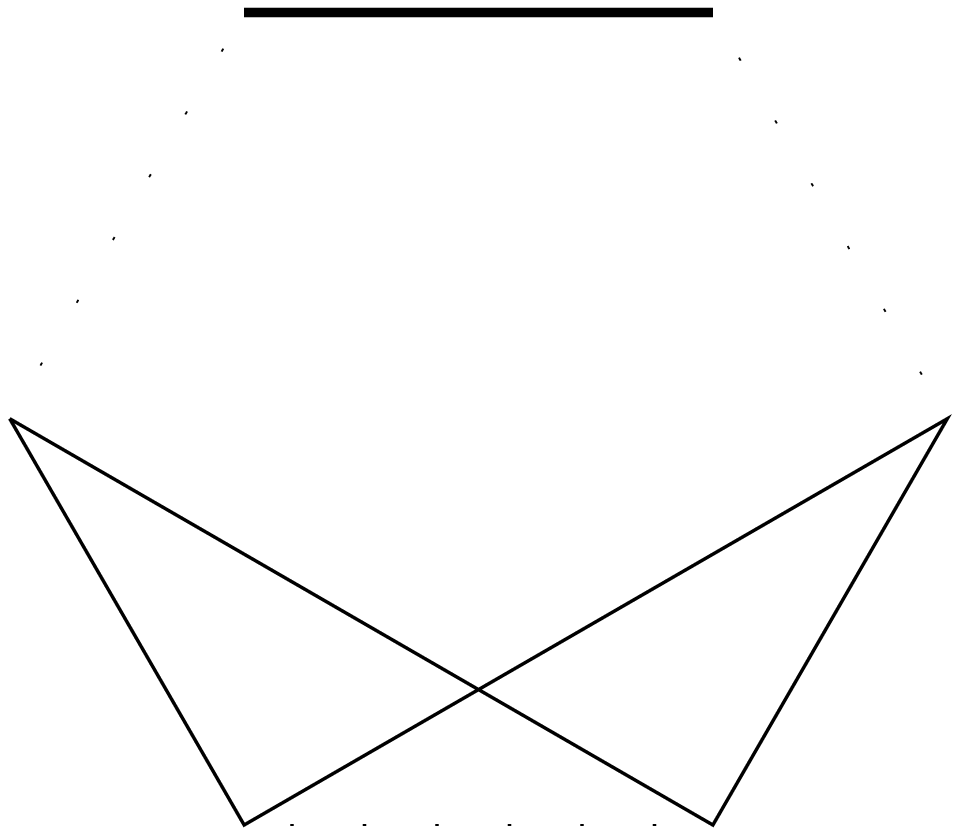}
\hfill
\includegraphics[angle=-90,scale=.2]{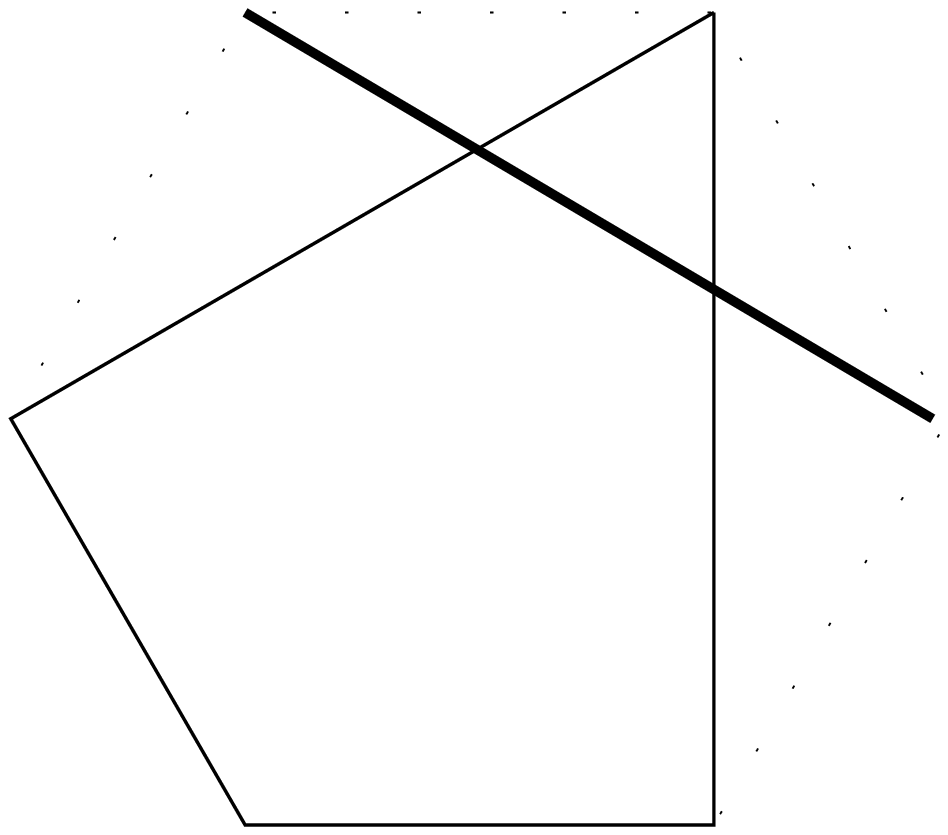}
\hfill
\includegraphics[angle=-90,scale=.2]{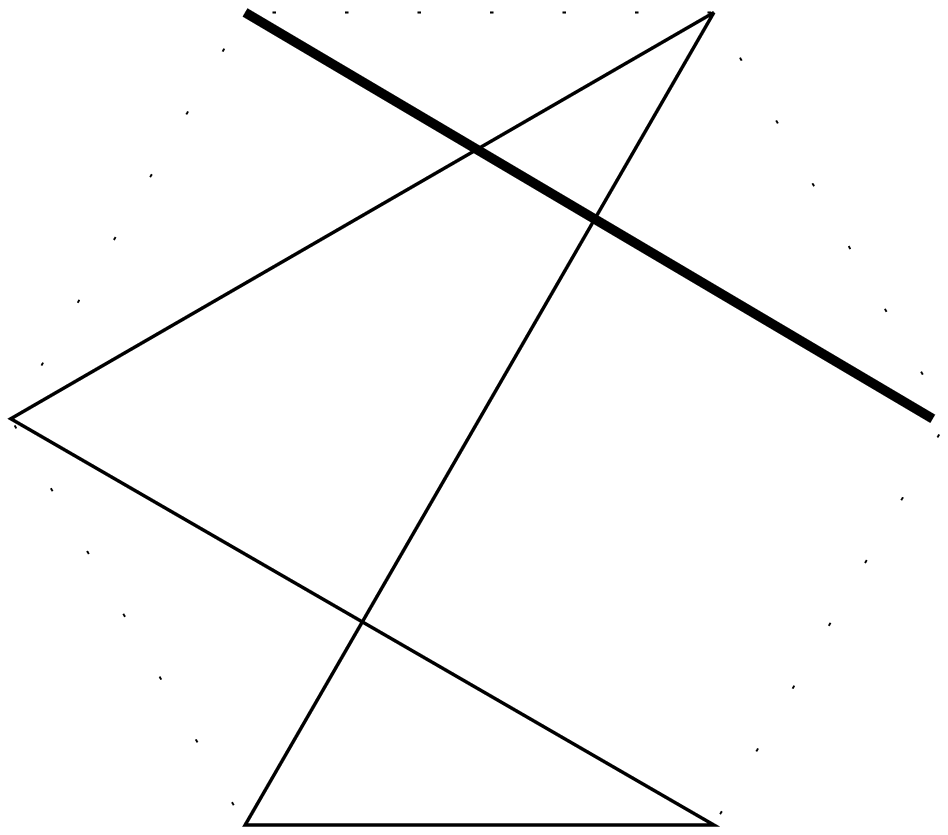}
\hfill \raisebox{-.07\textwidth}{5}
\\[1mm]

\hfill \hfill \raisebox{-.07\textwidth}{6} \hfill
\includegraphics[angle=-90,scale=.2]{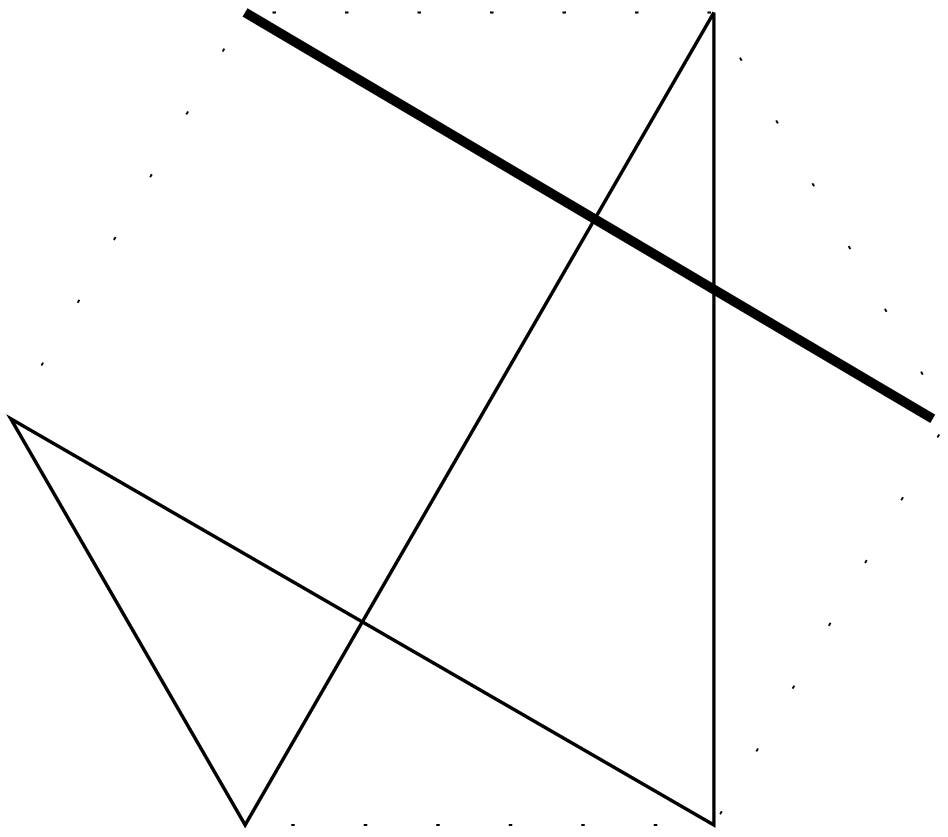}
\hfill
\includegraphics[angle=-90,scale=.2]{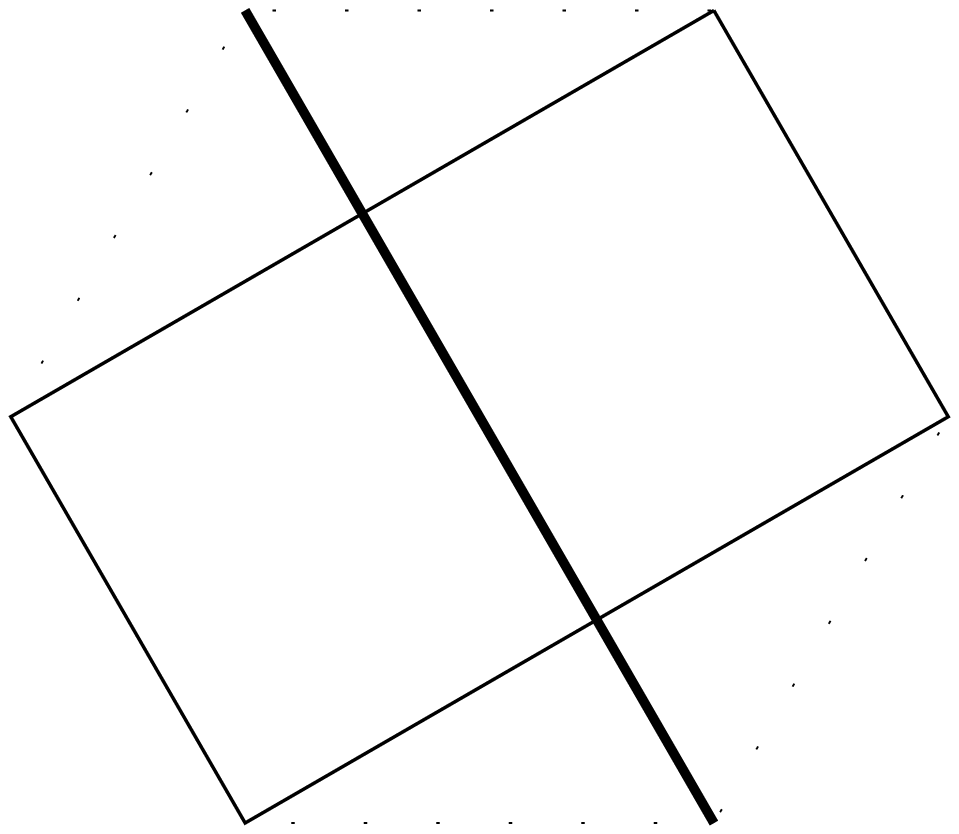}
\hfill
\includegraphics[angle=-90,scale=.2]{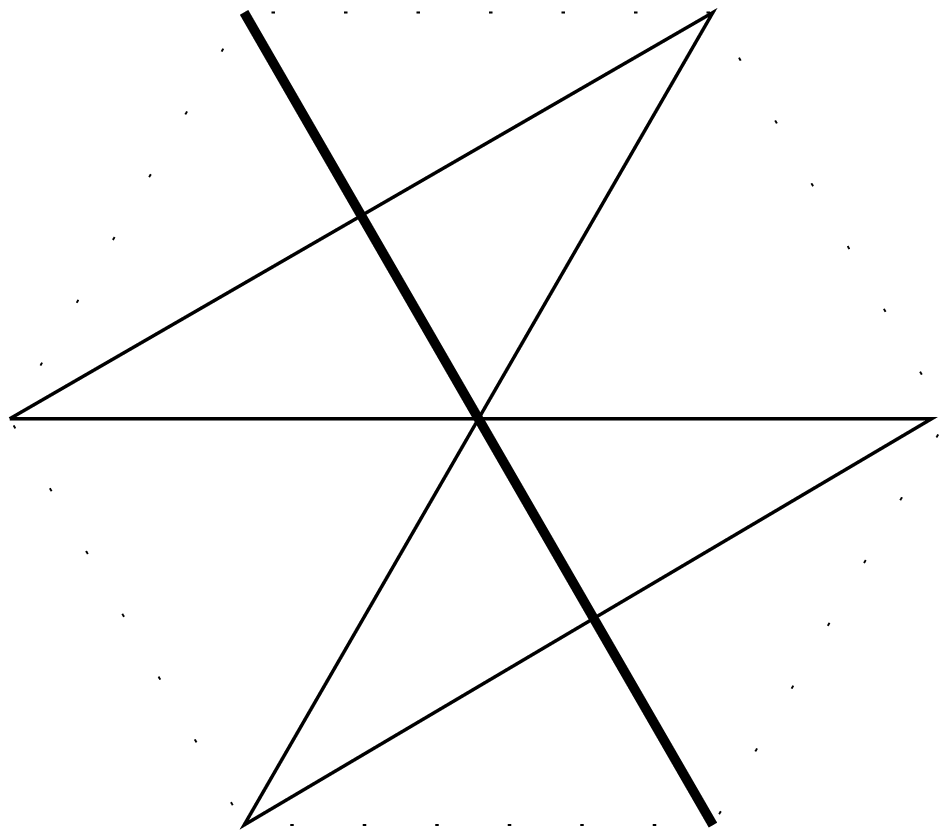}
\hfill
\includegraphics[angle=-90,scale=.2]{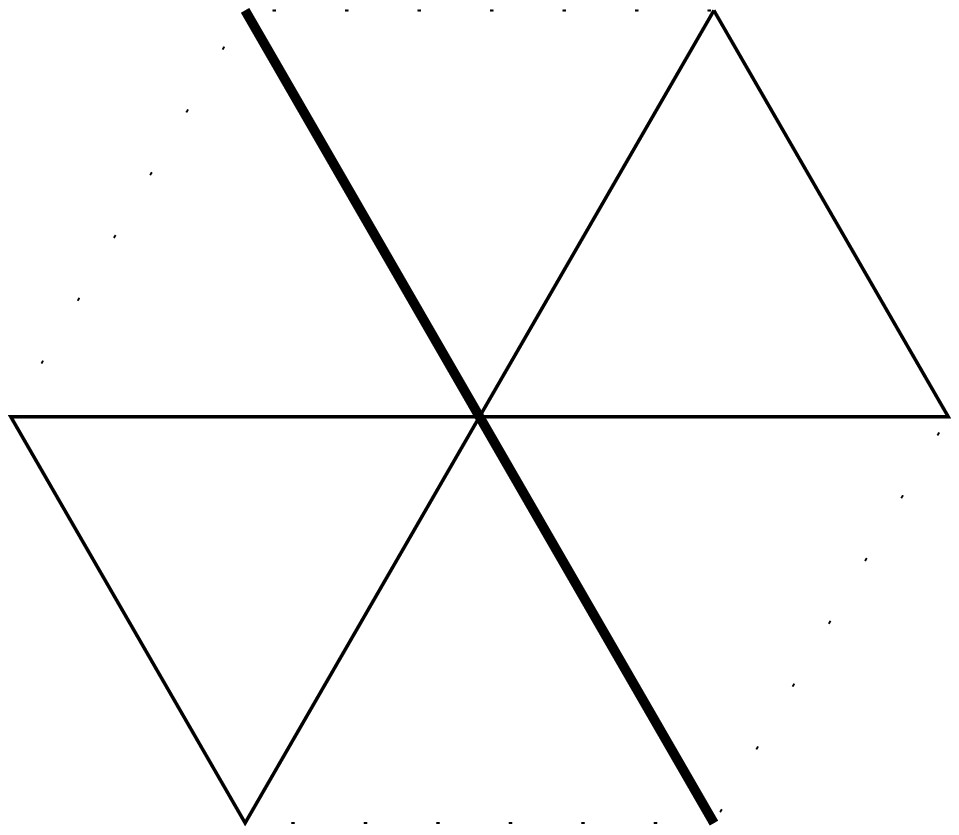}
\hfill \raisebox{-.07\textwidth}{9}
\hfill
\end{minipage}
\caption{Diagrammatic  representation of the nine [42] terms in the action,
numbered as indicated, from left to right.} 
\label{negen keer 42}
\end{figure}

\begin{figure}[h]
\begin{minipage}[t]{\textwidth}
\hfill \raisebox{-.07\textwidth}{1}
\hfill
\includegraphics[angle=-90,scale=.2]{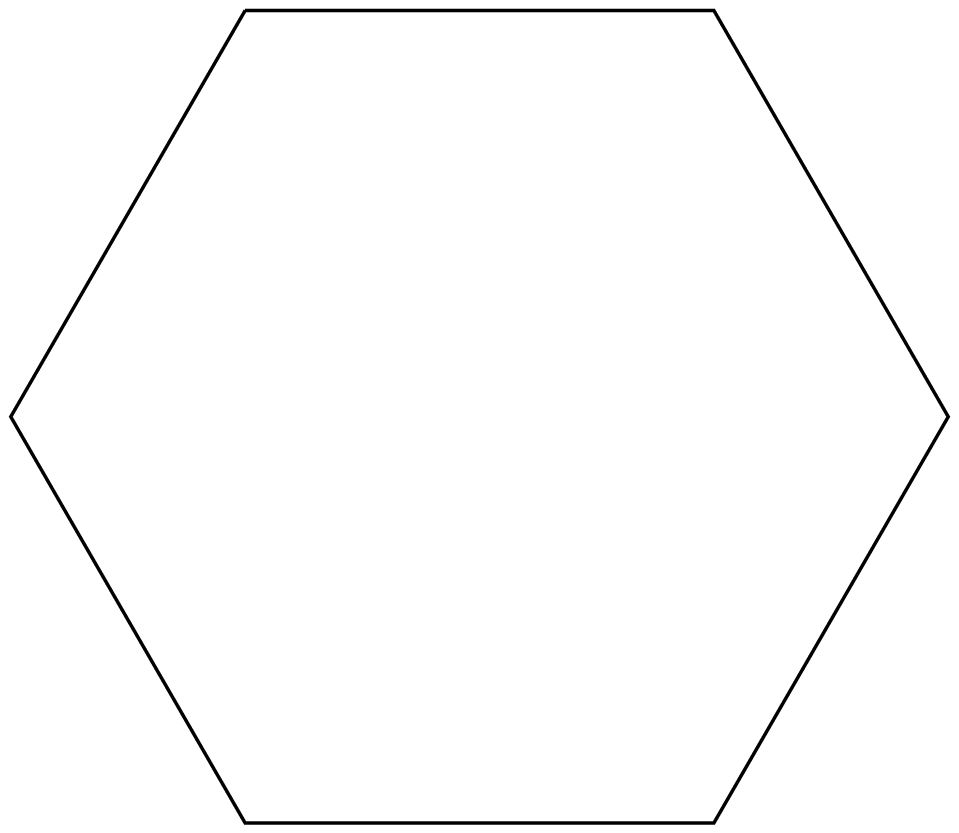}
\hfill
\includegraphics[angle=-90,scale=.2]{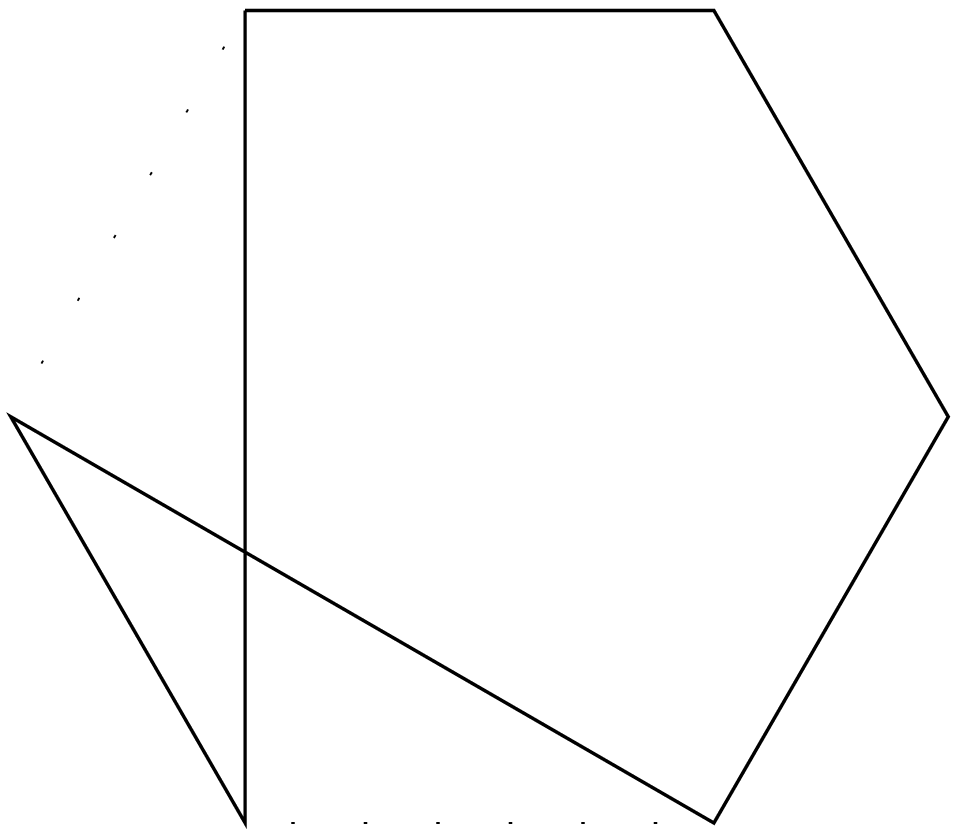}
\hfill
\includegraphics[angle=-90,scale=.2]{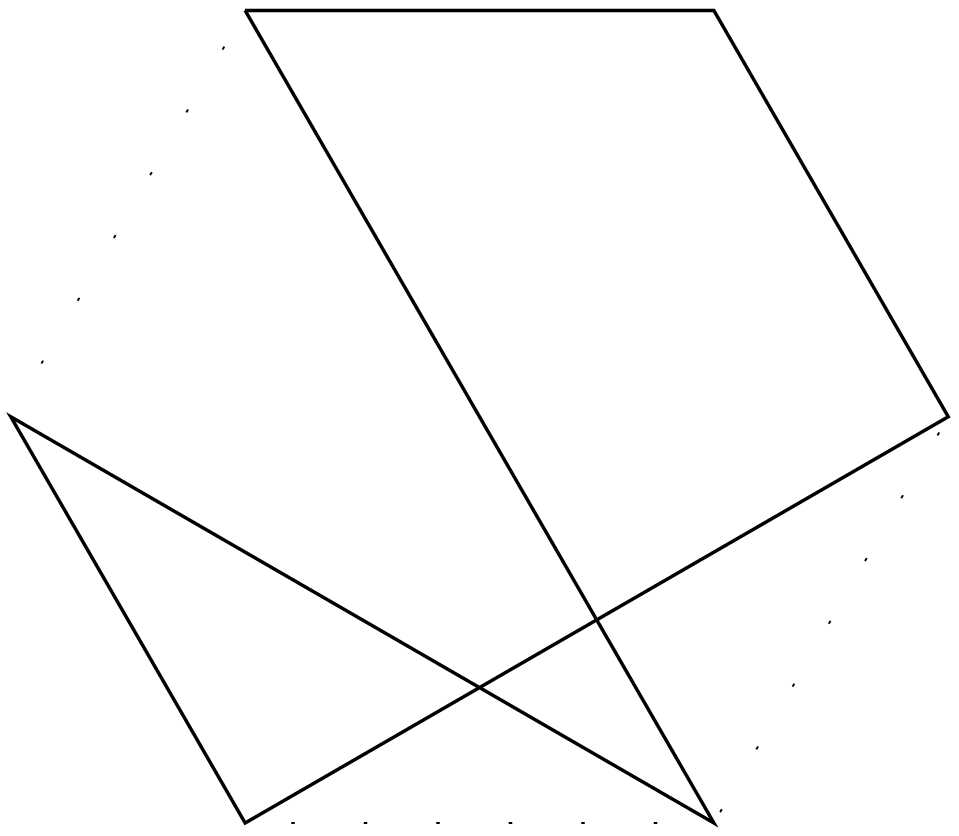}
\hfill
\includegraphics[angle=-90,scale=.2]{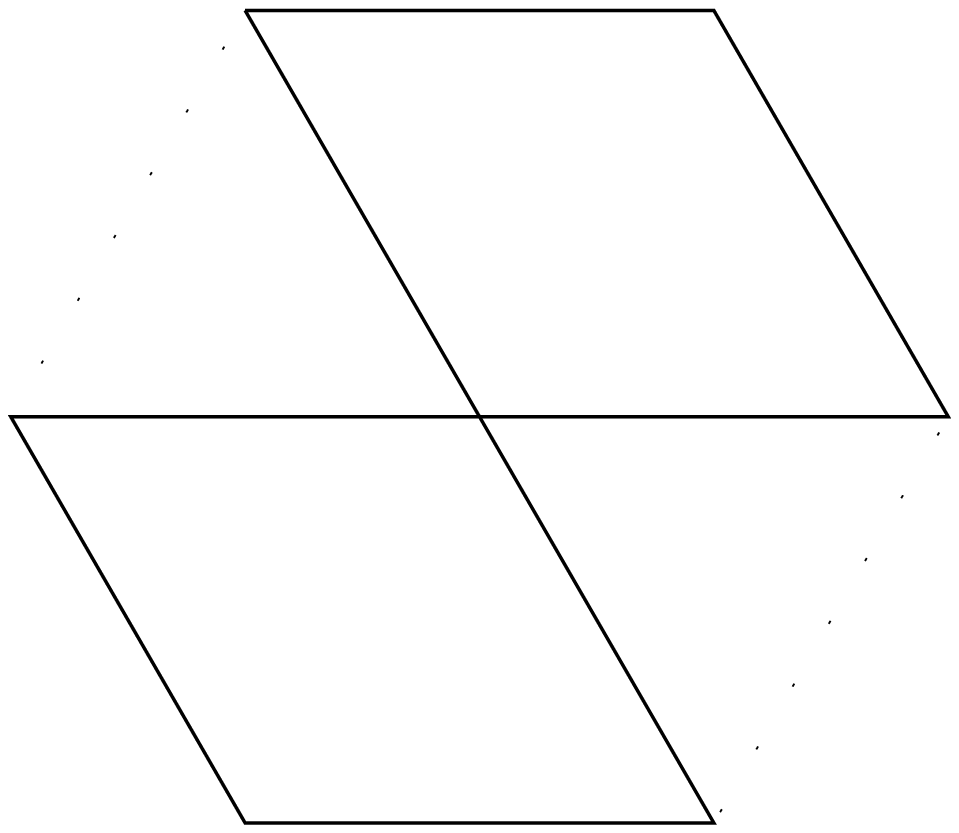}
\hfill
\includegraphics[angle=-90,scale=.2]{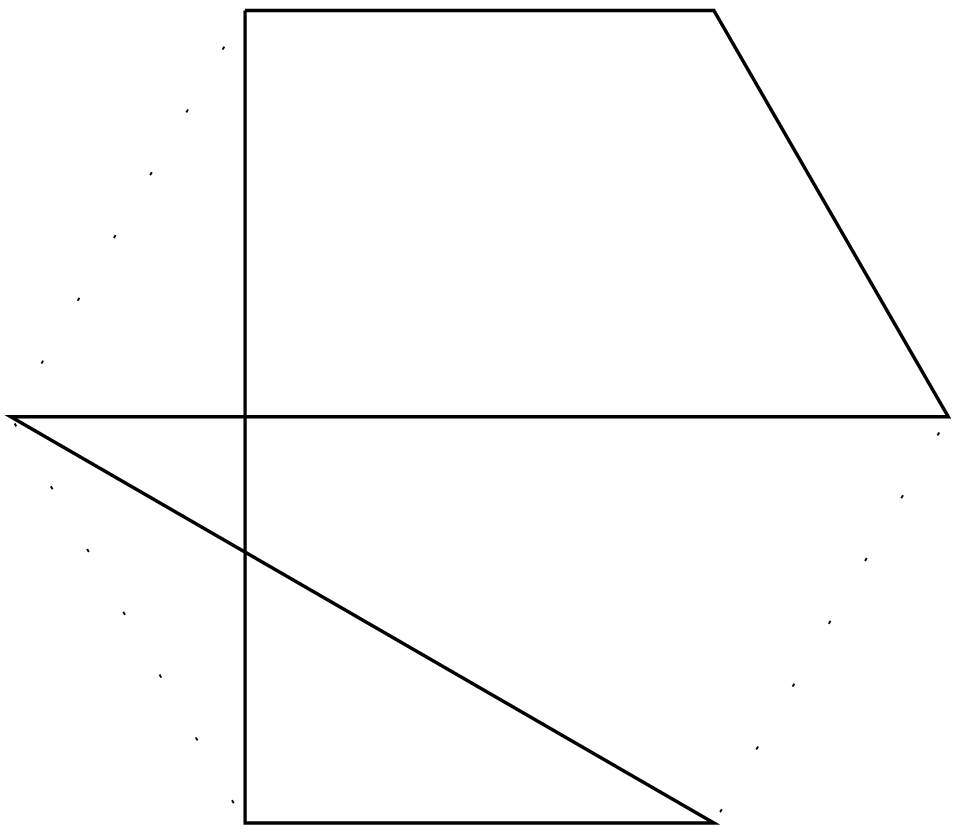}
\hfill \raisebox{-.07\textwidth}{5}
\\[1mm]

\hfill \hfill \raisebox{-.07\textwidth}{6} \hfill
\includegraphics[angle=-90,scale=.2]{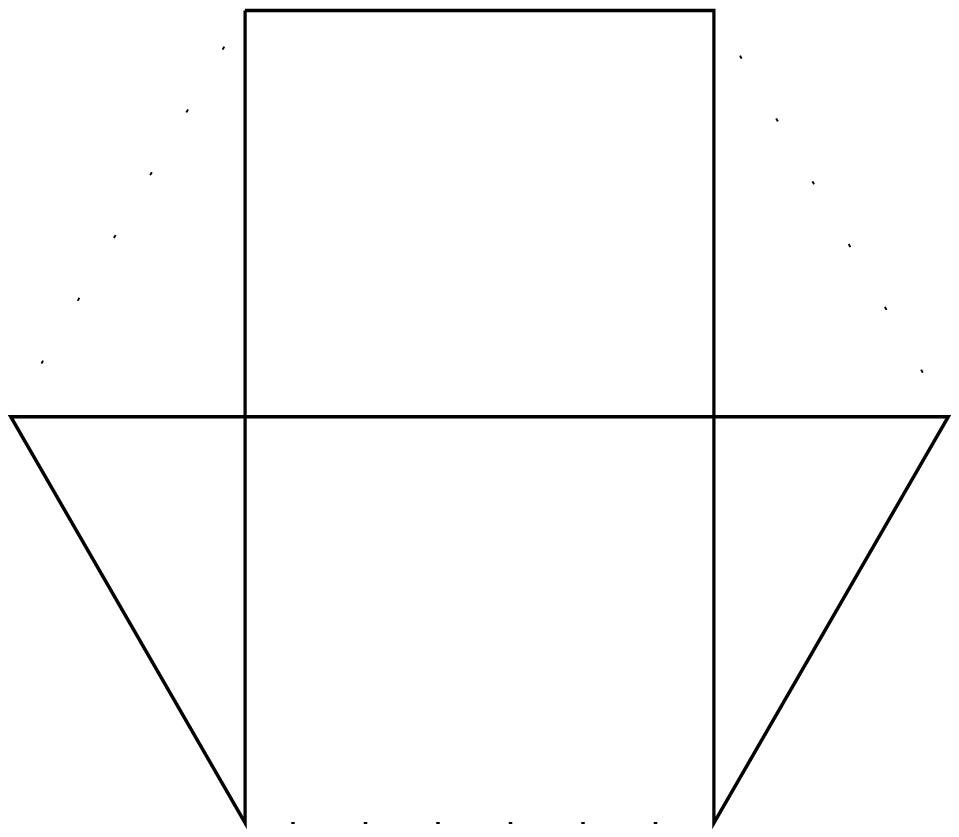}
\hfill
\includegraphics[angle=-90,scale=.2]{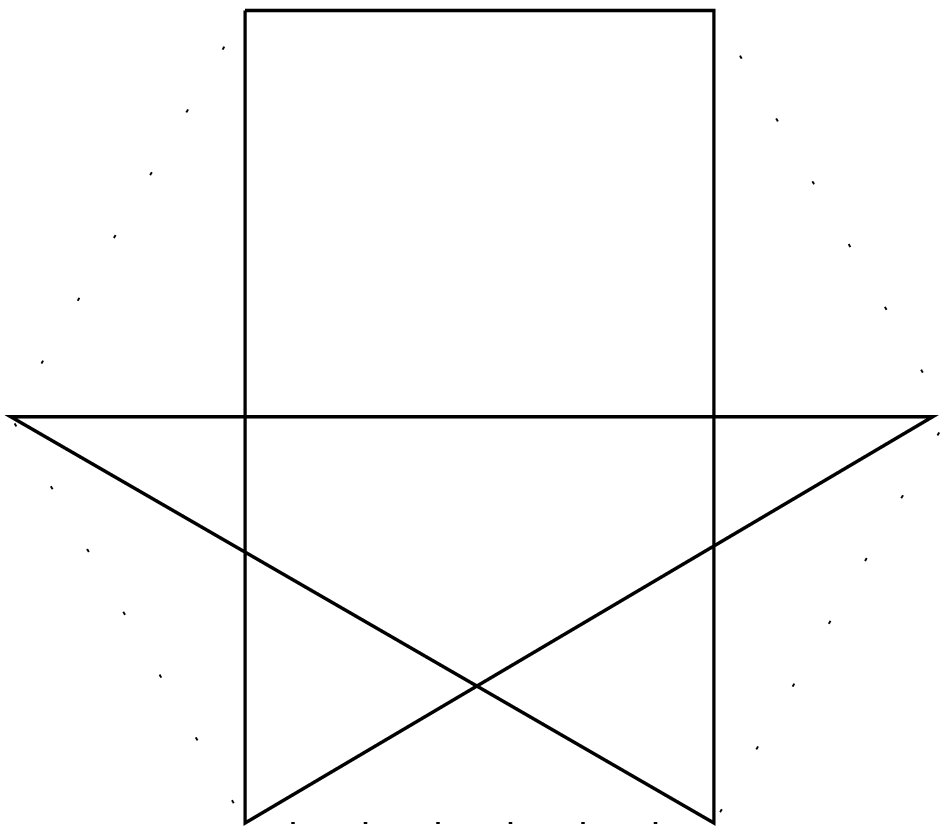}
\hfill
\includegraphics[angle=-90,scale=.2]{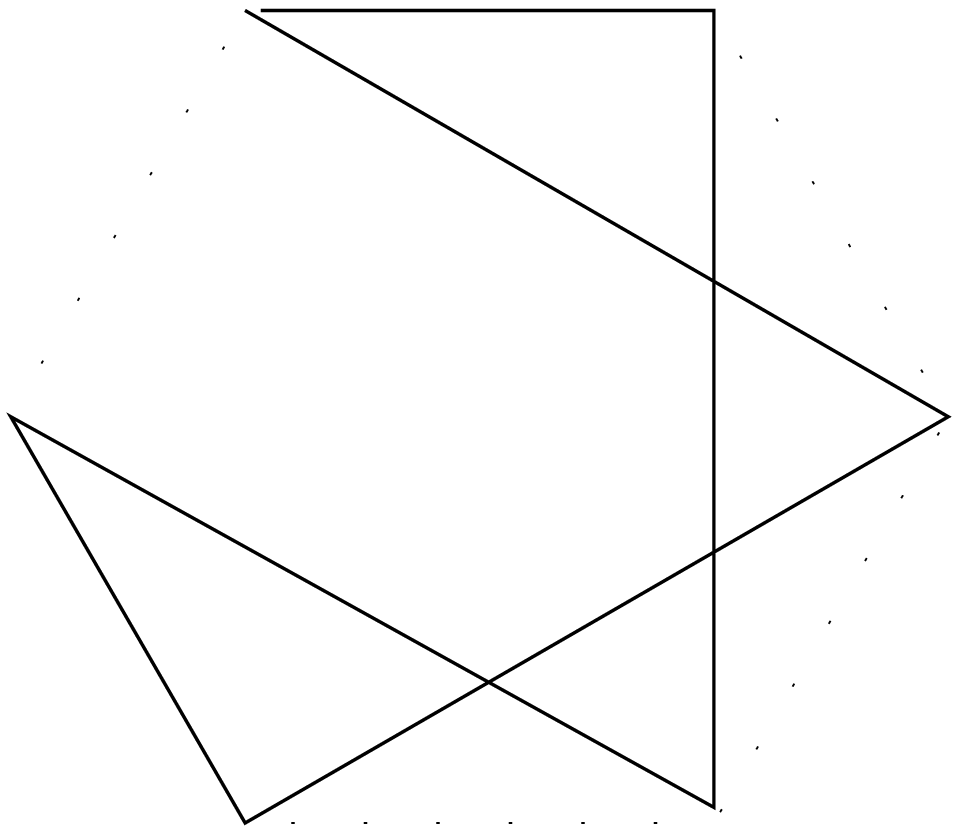}
\hfill
\includegraphics[angle=-90,scale=.2]{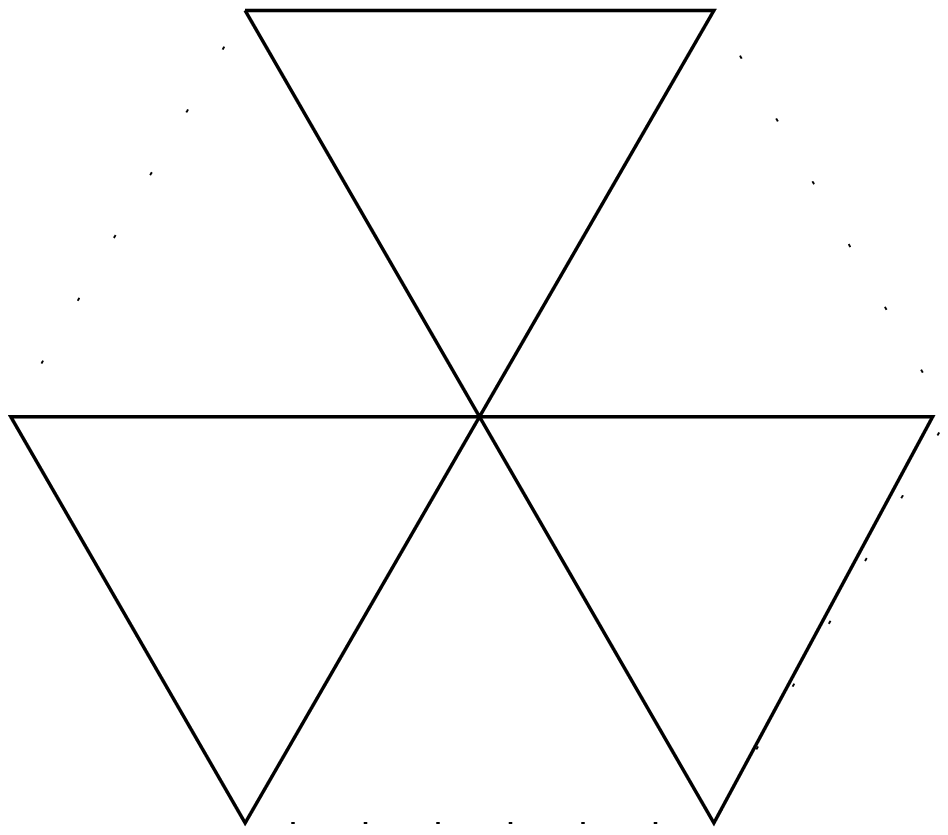}
\hfill
\includegraphics[angle=-90,scale=.2]{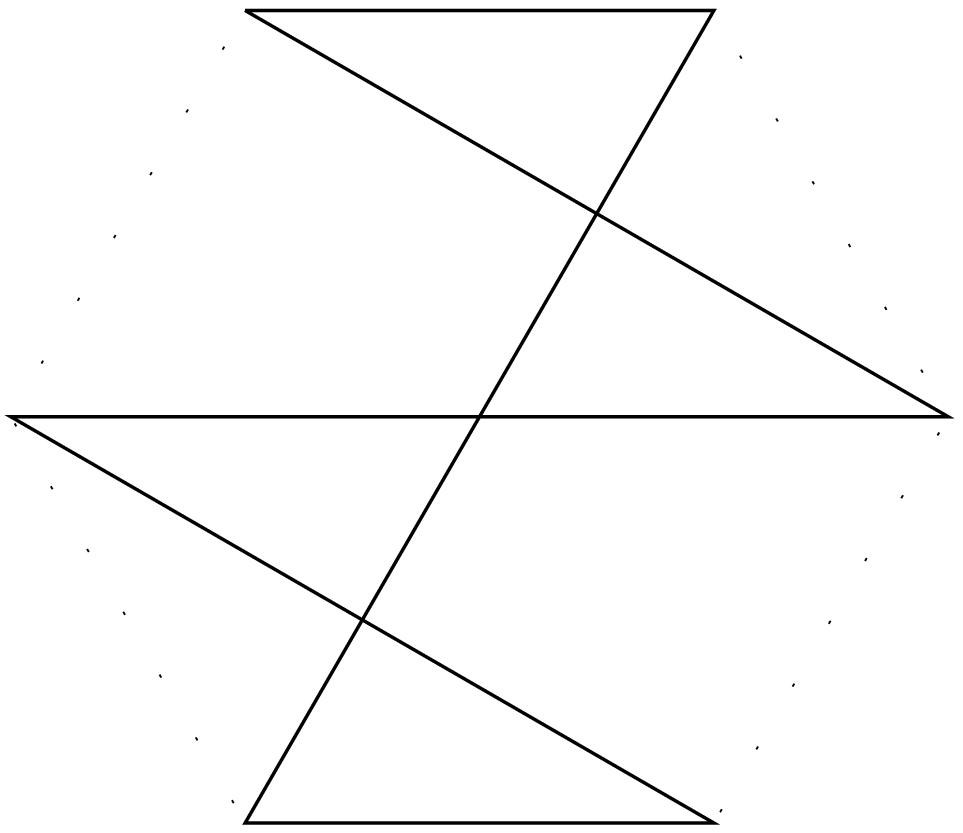}
\hfill \raisebox{-.07\textwidth}{10}
\hfill
\\[1mm]

\hfill \hfill \raisebox{-.07\textwidth}{11} \hfill
\includegraphics[angle=-90,scale=.2]{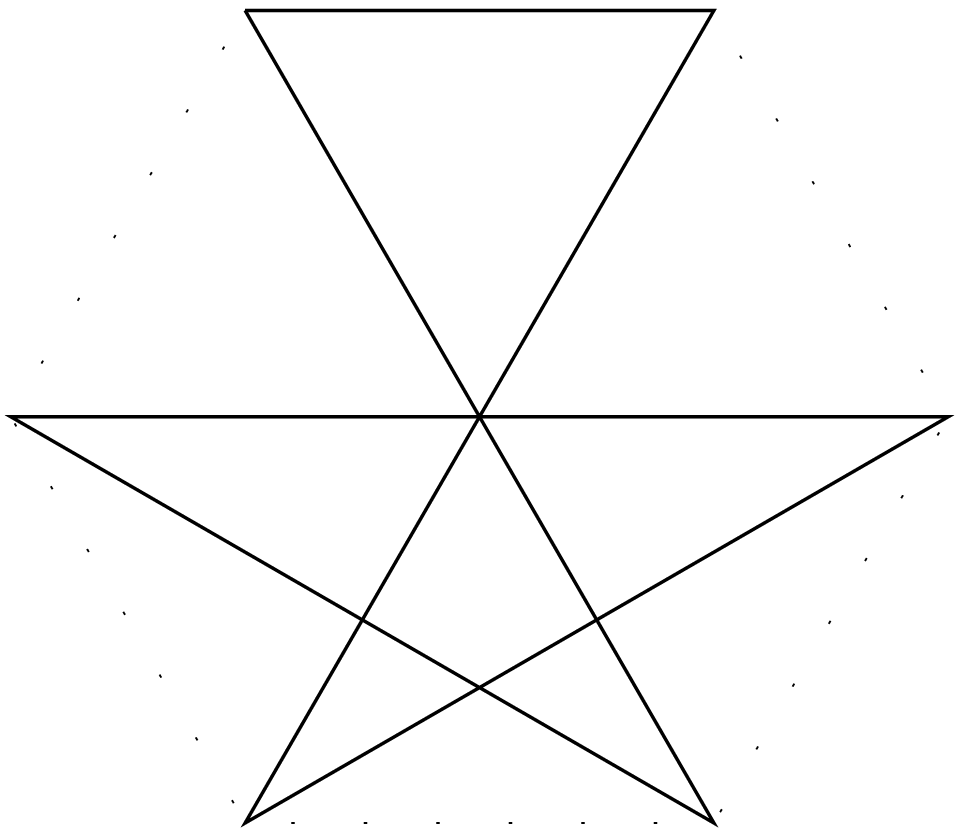}
\hfill
\includegraphics[angle=-90,scale=.2]{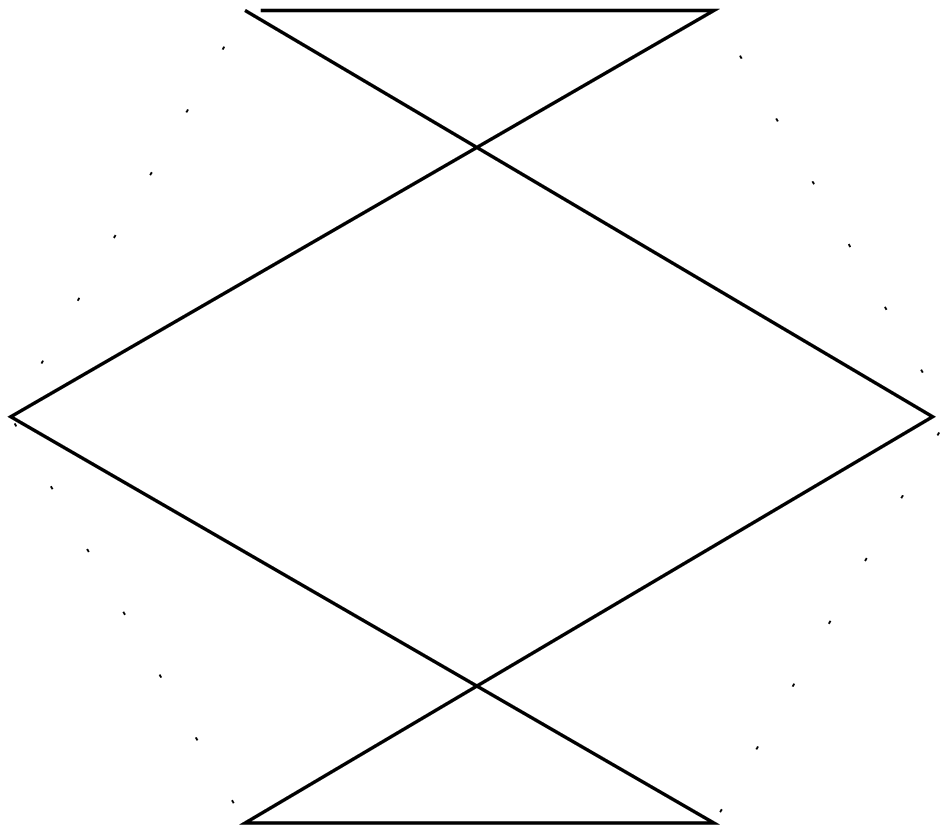}
\hfill
\includegraphics[angle=-90,scale=.2]{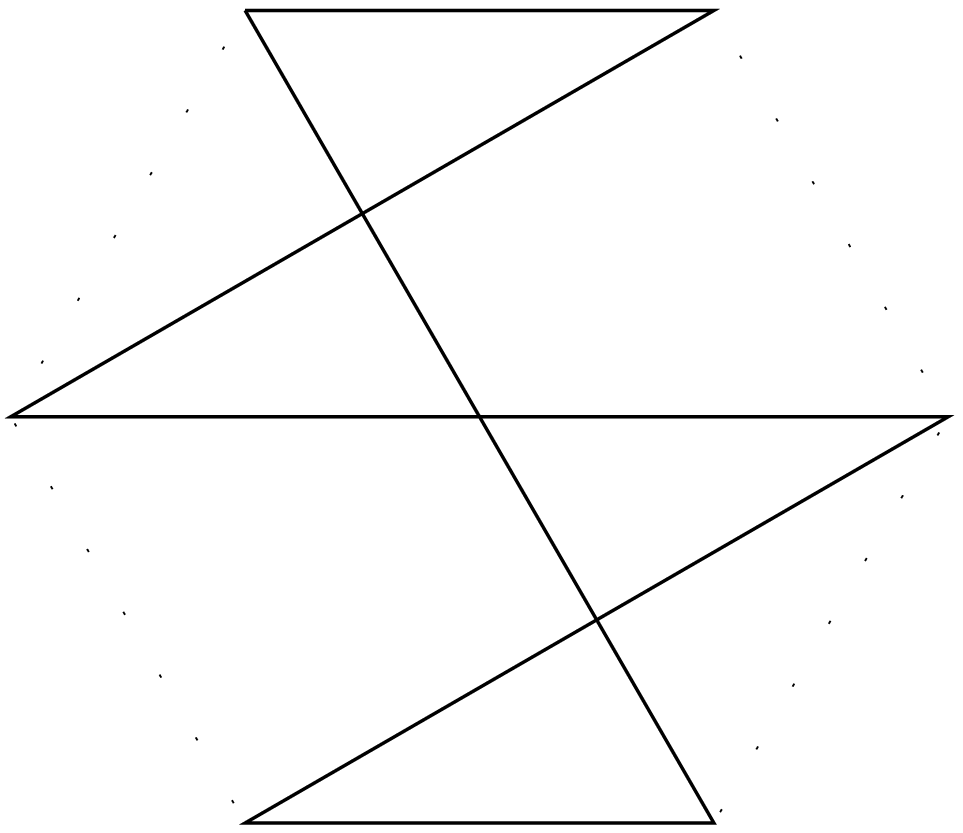}
\hfill
\includegraphics[angle=-90,scale=.2]{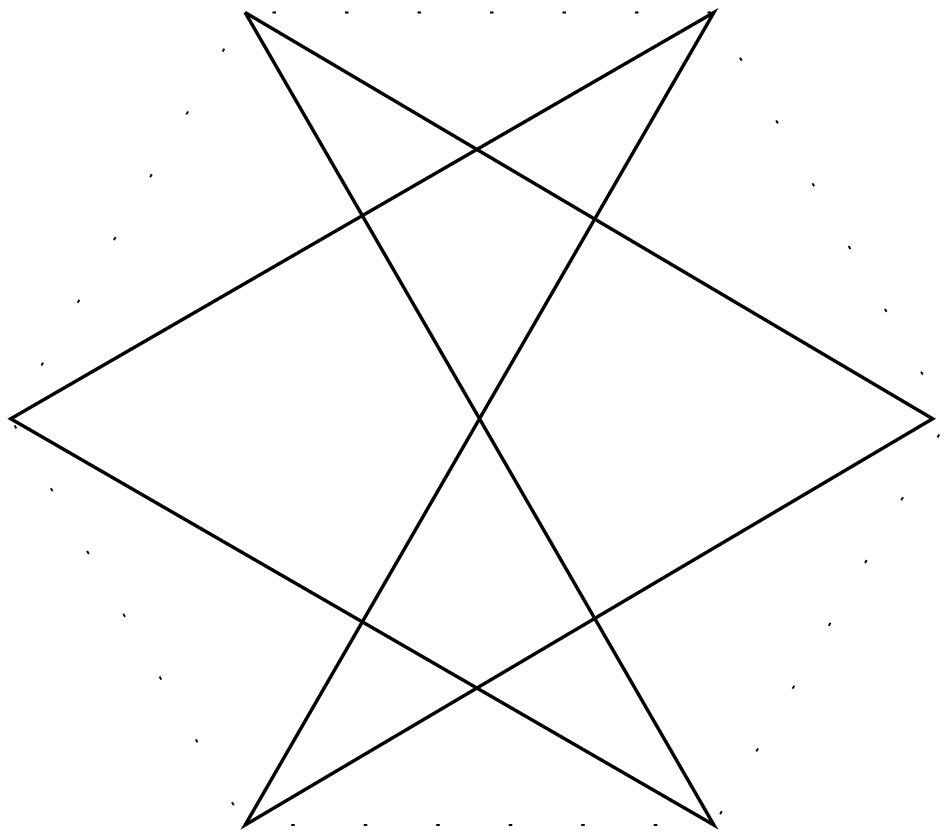}
\hfill \raisebox{-.07\textwidth}{14}
\hfill
\end{minipage}
\caption{Diagrammatic  representation of the fourteen [6] terms in the action.} 
\label{veertien keer 6}
\end{figure}

 In (\ref{genact}) we give a few examples in 
our ansatz for the action that should leave no ambiguity as to which terms 
in the action the diagrams correspond to. The upper index on $a^{nnn}_m$
indicates the class, the lower index the number of the diagram:
\begin{eqnarray}
{\cal L}= Tr ( \dots 
\nonumber \\
&+& \dots +  a^6_3  F_{\alpha_1 \alpha_2} F_{\alpha_2 \alpha_3} F_{\alpha_3 \alpha_4} F_{\alpha_6 \alpha_1}
F_{\alpha_4 \alpha_5} F_{\alpha_5 \alpha_6}
\nonumber \\
&+&\dots + a^6_4  F_{\alpha_1 \alpha_2} F_{\alpha_2 \alpha_3} F_{\alpha_3 \alpha_4} F_{\alpha_6 \alpha_1}
F_{\alpha_5 \alpha_6} F_{\alpha_4 \alpha_5} 
\nonumber \\ 
&+&\dots  + a^6_{14}  F_{\alpha_1 \alpha_2} F_{\alpha_4 \alpha_5} F_{\alpha_2 \alpha_3} F_{\alpha_6 \alpha_1}
F_{\alpha_3 \alpha_4} F_{\alpha_5 \alpha_6}
\nonumber \\
&+&\cdots + a^{2,4}_1  F_{\alpha_1 \alpha_2} F_{\alpha_2 \alpha_1} F_{\beta_1 \beta_2} F_{\beta_2 \beta_3}
F_{\beta_3 \beta_4} F_{\beta_4 \beta_1}
\nonumber \\
&+&\cdots + a^{2,2,2}_3  F_{\alpha_1 \alpha_2}  F_{\beta_1 \beta_2} F_{\gamma_1 \gamma_2} F_{\alpha_2 \alpha_1}
F_{\beta_2 \beta_1}   F_{\gamma_2 \gamma_1}  \nonumber \\
&+&    O(F^8)     \,.        \label{genact}
\end{eqnarray}
\section{Table}\label{BigTable}
\begin{center}
\begin{table}
\begin{tabular}{|c|c|c|c|c|}
\hline
Class & $S_6$-rep & Prefactor & Invariant linear combination  \\
\hline
 2 2 2 & $ [6]  $ & $ \frac{1}{15} $ & $  3 i_1 + 2 i_2 
                              +  i_3 + 3 i_4 + 6 i_5 $ \\  \hline
 2 2 2 & $ [4 2] $ & $ \frac{1}{10} $ & $ 2  i_1 +3 i_2 
                              - i_3 -3  i_4 - i_5 $   \\ \hline
 2 2 2 & $ [4 2] $ & $ \frac{1}{20} $ & $  -3 i_1-2 i_2
                               -  i_3+ 2 i_4 + 4 i_5 $ \\  \hline  
 2 2 2 & $ [2^3]  $ & $ \frac{1}{12} $ & $  -6 i_1 + 5 i_2 
                              +   i_3 + 3 i_4 - 3 i_5 $ \\  \hline
 2 2 2 & $ [2^3] $ & $ \frac{1}{12} $ & $  0 i_1-  i_2
                              +  i_3- 3 i_4 +3 i_5 $  \\ \hline 
 4 2 & $ [6] $ & $ \frac{1}{15} $ & $2,2,2,2,2,2,1,1,1$ \\  \hline  
 4 2 & $ [4 2] $ & $ \frac{1}{10} $ & $3,-3,0,2,-1,-1,1,-2,1$ \\  \hline  
 4 2 & $ [4 2] $ & $ \frac{1}{20} $ & $5,3,4,-2,-3,-3,-1,-2,-1$ \\  \hline  
 4 2 & $ [4 2] $ & $ \frac{1}{20} $ & $-4,2,2,0,0,0,1,1,-2$ \\  \hline  
 4 2 & $ [4 2] $ & $ \frac{1}{20} $ & $-4,-2,-2,4,4,4,-1,-1,-2$\\  \hline  
 4 2 & $ [3 2 1] $ & $ \frac{1}{15} $ & $-1,0,1,-1,-2,3,0,-1,1$\\  \hline  
 4 2 & $ [3 2 1] $ & $ \frac{1}{15} $ & $-1,0,1,-1,3,-2,0,-1,1$ \\  \hline  
 4 2 & $ [2^3] $ & $ \frac{1}{12} $ & $1,-5,4,-2,1,1,-1,2,-1$ \\  \hline  
 4 2 & $ [2^3] $ & $ \frac{1}{6} $ &  $1,1,-2,-2,1,1,2,-1,-1$ \\  \hline  
 6 & $[6]  $ & $  \frac{1}{60} $ & $ 1,6,6,3,6,6,6,6,2,3,6,3,3,3$  \\ \hline  
 6 & $[4 2]   $ & $ \frac{1}{20} $ & $1,-1,-1,3,-1,-1,0,0,2,-1/2,-1,0,-1/2,0$\\ \hline  
 6 & $[4 2]   $ & $ \frac{1}{40} $ & $1,4,1,0,1,1,-5,-2,-1,1,-1,1,1,-2 $\\ \hline    
 6 & $[4 2]   $ & $ \frac{1}{20} $ & $1,2,-1,0,-1,-1,-1,2,-1,0,-3,3,0,0 $\\ \hline                                                      
 6 & $[4 2]   $ & $ \frac{1}{20} $ & $1,2,1,2,1,1,-1,0,1,-1,-3,-1,-1,-2 $\\ \hline
 6 & $[3 2 1]   $ & $ \frac{1}{45} $ & $4,4,2,0,2,-8,4,-4,-4,-2,4,-4,-2,4 $\\ \hline
 6 & $[3 2 1]   $ & $ \frac{1}{45} $ & $1,1,3,0,-2,-2,1,-1,-1,-3,1,-1,2,1 $\\ \hline 
 6 & $[2^3]   $ & $ \frac{1}{12} $ & $ 1,-4,2,-1,2,2,2,-4,0,-1,0,3,-1,-1$\\ \hline
 6 & $[2^3]   $ & $ \frac{1}{36} $ & $1,-2,-1,0,-1,7,1,-4,-1,1,-5,-1,1,4$\\ \hline                                                     
 6 & $[2^3]  1 $ & $ \frac{1}{12} $ & $1,3,-3,-3,-3,3,3,-3,2,0,0,0,0,0$\\ \hline
 6 & $[2^3]   $ & $ \frac{1}{24} $ & $1,2,-3,-2,-3,5,-1,0,1,-1,1,-1,-1,2$\\ \hline
 6 & $[3 1^3]   $ & $ \frac{1}{18} $ & $0,0,-1,0,1,0,0,0,0,-1/2,0,0,1/2,0$\\ \hline 
 6 & $[3 1^3]   $ & $ \frac{1}{18} $ & $1,-2,-3,0,1,1,1,2,-1,0,1,-1,2,-2 $\\ \hline                                                      
 6 & $[2 1^4]   $ & $ \frac{1}{36} $ & $ 1,-2,2,-3,2,-2,-2,2,2,1,-2,-1,1,1$\\ \hline
\end{tabular}
\caption{Cyclic invariants by irreducible representation} 
\label{invariants}
\end{table}
\end{center}
The table in this appendix records the result of the construction of a basis of 
invariants based on permutation group analysis. The notation is as follows.

The first five lines correspond to the class of terms with three times a double contraction
of Lorentz indices, labeled class 222. The second column gives the
permutation group class in the standard cycle notation, and the third gives the corresponding
invariant. The combination is written as a weighted sum of diagrams $i_n$,
the latter labeled in the order given in figure~\ref{vijf keer 
222} (see appendix~\ref{diagrams}). In the following lines, 
this information is given for the classes of terms with Lorentz contractions following
the patterns $4 2$ and $6$ respectively. For the invariant linear combinations
we just give the coefficients,
again corresponding to the figures (\ref{negen keer 42}~and~\ref{veertien keer 6}) in 
the preceding appendix.

\pagebreak
\section{Abelian constraint}
\label{abelianconstr}
We know the Born-Infeld action for the gauge group $U(1)$.
After expanding the determinant and the square root it looks as follows:
\begin{eqnarray}
{\cal L}^6&=& \frac{1}{4} F^{\alpha \beta} F_{\beta \alpha} \nonumber \\
& & +\frac{1}{8} F^{\alpha \beta} F_{\beta \gamma} F^{\gamma \delta} 
F_{\delta \alpha} - \frac{1}{32} (F^{\alpha \beta} F_{\alpha \beta})^2 
\nonumber \\
& & +  \frac{1}{12} F^{\alpha \beta} F_{\beta \gamma} F^{\gamma \delta} 
F_{\epsilon \zeta} F^{\zeta \iota} F_{\iota \alpha} -
  \frac{1}{32} F^{\alpha \beta} F_{\beta \alpha} F^{\gamma \delta} 
F_{\delta \zeta} F^{\zeta \iota} F_{\iota \gamma} \nonumber \\
& & 
+ \frac{1}{384} (F^{\alpha \beta} F_{\beta \alpha})^3 \nonumber  \\
& &   + O(F^8)\,.
\end{eqnarray}
{From} this we 
derive the following constraints on the coefficients in our general ansatz:
\begin{eqnarray}
a^2_1 & =& \frac{1}{4} \nonumber \,,\\
a_1^4+a_2^4 &=& \frac{1}{8} \nonumber\,, \\
a_1^{2,2}+a_2^{2,2} &=& -\frac{1}{32} \,, \label{abelian24}
\end{eqnarray}
up to fourth order, and
\begin{eqnarray}
\sum_{i=1}^{14} a_i^6 &=& \frac{1}{12} \,,\nonumber \\
\sum_{i=1}^9 a_i^{2,4} &=& -\frac{1}{32} \,,\nonumber \\
\sum_{i=1}^5 a_i^{2,2,2} &=& \frac{1}{384} \,,        \label{abelian6}
\end{eqnarray}
at sixth order.

\section{Some Technical details}
\label{appactquad}\label{notationsappendix}
\subsection*{Action quadratic in fluctuations \\
         in a magnetic background}
We split the field strength in background and fluctuations,
$F={\cal F} + \delta F$, and the fluctuations into a part linear in the 
gauge field fluctuations and a part quadratic in the gauge field 
fluctuations:
\begin{eqnarray}
F_{\alpha \beta} &=&  \partial_{\alpha} A_{\beta} -
\partial_{\beta} A_{\alpha} + i [ A_{\alpha}, A_{\beta} ] \,, \\
\delta F_{\alpha \beta} &=& \delta_1 F_{\alpha \beta} +    \delta_2 F_{\alpha \beta}\,, \\
\delta_1 F_{\alpha \beta} &=& D_{\alpha} \delta A_{\beta}
                              - D_{\beta} \delta A_{\alpha} \label{delta1}\,,\\
\delta_2 F_{\alpha \beta} &=&  i [ \delta A_{\alpha}, \delta A_{\beta} ] \,,\\
D_{\alpha} &=& \partial_{\alpha} + i [ A_{\alpha}, . ] \,.
\end{eqnarray}
We substitute $F={\cal F} + \delta_1 F+ \delta_2 F$  into the action up to 
order $F^6$ and restrict to the terms quadratic in the fluctuations. 
We'll get terms proportional to $\delta_2 F$ and terms proportional to 
$(\delta_1 F)^2$. 
\subsection*{Terms proportional to $\delta_2 F$}
When  we choose $ {\cal F} $ in the CSA we can write the 
terms proportional to $\delta_2 F$ as:
\begin{eqnarray}
{\cal L}^{(2)}_2 &=& \frac{1}{2} \delta_2 F^{\alpha \beta} 
{\cal F}_{\beta \alpha} \nonumber \\
& & +\frac{1}{2} \delta_2 F^{\alpha \beta} {\cal F}_{\beta \gamma} 
{\cal F}^{\gamma \delta} 
{\cal F}_{\delta \alpha} - \frac{1}{8} \delta_2 F^{\alpha \beta} {\cal F}_{\alpha \beta}
{\cal F}^{\delta \gamma} {\cal F}_{\delta \gamma}
\nonumber \\
& & +  \frac{1}{2} \delta_2 F^{\alpha \beta} {\cal F}_{\beta \gamma} 
{\cal F}^{\gamma \delta} 
{\cal F}_{\delta \zeta} {\cal F}^{\zeta \iota} {\cal F}_{\iota \alpha} -
  \frac{1}{16} \delta_2 F^{\alpha \beta} {\cal F}_{\beta \alpha}
  {\cal F}^{\gamma \delta} 
{\cal F}_{\delta \zeta} {\cal F}^{\zeta \iota} {\cal F}_{\iota \gamma} \nonumber \\
& & -
\frac{1}{8} \delta_2  F^{\gamma \delta} 
{\cal F}_{\delta \zeta} {\cal F}^{\zeta \iota} {\cal F}_{\iota \gamma}   
{\cal F}^{\alpha \beta} {\cal F}_{\beta \alpha}
+ \frac{1}{64} \delta_2 F_{\alpha \beta} {\cal F}^{\beta \alpha} 
({\cal F}^{\gamma \delta} {\cal F}_{\gamma \delta})^2 \nonumber\,.  \\
& &
\end{eqnarray} 
This part of the  action naturally has the same coefficients as the 
abelian action. 
\subsection*{Terms proportional to $(\delta_1 F)^2$}
The $U(2)$ components of $\delta_1F$ (see eq. (\ref{delta1}) are denoted as in
$\delta_1 F=\sum_{n=1,2}\delta_1F^{(n)}\sigma_n$ and the 
background splits likewise as ${\cal F}={\cal F}^0 + {\cal F}^3 \sigma_3$. For the Lorentz index 
contraction we use a shorthand notation indicating the sequence(s) of contractions,
easily understood and generalised from the following hypothetical example:
$$A_{\mu_1\mu_2} B_{\mu_3 \mu_1} C_{\nu_1\nu_2} D_{\mu_2 \mu_3} E_{\nu_2\nu_1}
\longrightarrow A_1 B_3 C_{1'} D_2 E_{2'} .$$
Our calculation to order four gives for the off-diagonal fluctuations:
\begin{eqnarray} 
\!\!\!\!\!\!\!\!\!{\cal L}^{(2)}_1 =&& 2 a_1^2 (\delta_1 F_1^1 \delta_1 F_2^1+
\delta_1 F_1^2 \delta_1 F_2^2) \nonumber \\
&+&
               2     (\delta_1 F_1^1 \delta_1 F_2^1+
\delta_1 F_1^2 \delta_1 F_2^2) [(4 a_1^4+4 a_2^4) {\cal F}_3^0 {\cal F}_4^0
                                              + 4 a_1^4{\cal F}_3^3 {\cal F}_4^3]
                                              \nonumber \\ 
&+& 2 (\delta_1 F_1^1 \delta_1 F_3^1+
\delta_1 F_1^2 \delta_1 F_3^2)[ (2a_1^4+2a_2^4) {\cal F}_2^0 {\cal F}_4^0
                                 + (-2a_1^4+2a_2^4)  {\cal F}_2^3 {\cal F}_4^3]\nonumber \\ 
&+&  2     (\delta_1 F_1^1 \delta_1 F_2^1+
\delta_1 F_1^2 \delta_1 F_2^2)\times
\nonumber \\ & &\;\;\; 
 [(2 a_1^{2,2}+2 a_2^{2,2}) {\cal F}_{1'}^0 {\cal F}_{2'}^0 
                                              + (2 a_1^{2,2} -2 a_2^{2,2})
                                              {\cal F}_{1'}^3 {\cal F}_{2'}^3] 
 \\ 
&+& 2 (\delta_1 F_1^1 \delta_1 F_{1'}^1+
\delta_1 F_1^2 \delta_1 F_{1'}^2)[ (4 a_1^{2,2}+4 a_2^{2,2}) {\cal F}_2^0 {\cal F}_{2'}^0
                                              + (4 a_2^{2,2})
                                              {\cal F}_2^3 {\cal F}_{2'}^3]\,\,.
\nonumber
\end{eqnarray} 
The corresponding expressions for the general form (with our restrictions)
of the action at order $F^6$ are not very illuminating, and we refrain from
giving them explicitely.

\subsection*{Background blockdiagonal in Lorentz indices}

After filling in the background we obtain, from the quadratic terms:
\begin{eqnarray}
\!\!\!\!\!\!\!\!\!{\cal L}^{(2)} 
& = & (\delta_1 F_{0, 2i-1}^{(a)})^2 + (\delta_1 F_{0,2i}^{(a)})^2 \nonumber \\
& & -   (\delta_1 F_{2i-1, 2i}^{(a)})^2 \nonumber \\
& & - \frac{1}{2} \sum_{i \neq j} (
(\delta_1 F_{2i-1, 2j-1}^{(a)})^2 + (\delta_1 F_{2i-1, 2j}^{(a)})^2
+(\delta_1 F_{2i, 2j-1}^{(a)})^2+(\delta_1 F_{2i, 2j}^{(a)})^2 ) \nonumber 
\\
& & -2  \delta_2 F_{2i-1,2i}^{(3)} f_i^3,   \label{bd2}
\end{eqnarray}
and from the quartic terms we find:
\begin{eqnarray}
\!\!\!\!\!\!\!\!\!\!\!{\cal L}^{(4)} 
& = & -8 [(\delta_1 F_{0, 2i-1}^{(a)})^2 + (\delta_1 F_{0,2i}^{(a)})^2][
 ( a_1^4+ a_2^4) (f_i^0)^2 +  a_1^4 (f_i^3)^2 
 \nonumber \\ & &  \;\;\;\;\;\;\;\;\;\;\;\;\;\;\;\;\;\;\;
  + (2 a_1^{2,2}+ 2 a_2^{2,2}) (f_k^0)^2+ (2 a_1^{2,2}-2 a_2^{2,2}) (f_k^3)^2
  ]
  \nonumber \\
& &+ 8   (\delta_1 F_{2i-1, 2i}^{(a)})^2
[ (3 a_1^4+3 a_2^4+4 a_1^{2,2} +4 a_2^{2,2}) (f_i^0)^2 
  \nonumber \\ & & \mbox{\hspace*{4cm}}
  + ( a_1^4+ a_2^4+4 a_2^{2,2}) (f_i^3)^2
  \nonumber \\ & & \mbox{\hspace*{4cm}}
  + (2 a_1^4+2 a_2^4) (f_k^0)^2 
  + (2 a_1^{2,2}-2 a_2^{2,2}) (f_k^3)^2 
  ]\nonumber \\
& &+ 8 \sum_{i \neq j} [
(\delta_1 F_{2i-1, 2j-1}^{(a)})^2 + (\delta_1 F_{2i-1, 2j}^{(a)})^2
+  (\delta_1 F_{2i, 2j-1}^{(a)})^2+(\delta_1 F_{2i, 2j}^{(a)})^2 ] \times
  \nonumber \\ & & \;\;\;\;\;\;
\times  [ ( a_1^4+ a_2^4) (f_i^0)^2 
  + ( a_1^4) (f_i^3)^2   
  + ( a_1^{2,2}+ a_2^{2,2}) (f_k^0)^2 
  + ( a_1^{2,2}- a_2^{2,2}) (f_k^3)^2 
  ] \nonumber \\ & & 
+8 (\delta_1 F_{2i-1, 2i}^{(a)} \delta_1 F_{2j-1, 2j}^{(a)})
  [  ( a_1^4+ a_2^4+4 a_1^{2,2}+4 a_2^{2,2}) f_i^0 f_j^0
   \nonumber \\ & &\mbox{\hspace*{5cm}}
 + ( -a_1^4+ a_2^4+4 a_2^{2,2}) f_i^3 f_j^3 
     ]   \nonumber \\ & & 
 +    \delta_2 F_{2i-1,2i}^{(3)} f_i^3
 [6 (f_i^0)^2+ 2(f_i^3)^2- (f_k^0)^2 -(f_k^3)^2]
         \nonumber \\ & & 
 +    \delta_2 F_{2i-1,2i}^{(3)} f_i^0
 (-2 f_k^0 f_k^3 )
     \nonumber 
\\
& & +8  \sum_{i \neq j} \delta_2 F_{2i-1,2i}^{(3)} f_i^3
[  ( a_1^4+ a_2^4) f_i^0 f_j^0
  + ( -a_1^4+ a_2^4) f_i^3 f_j^3 
     ]\,.                                            \label{bd4}
\end{eqnarray}
The sixth order calculation is analogous, the results are omitted.


\begin{thebibliography}{99}
\bibitem{T}  D. J. Gross and E. Witten, Nucl. Phys. {\bf B277} (1986) 1;
A.A. Tseytlin, Nucl. Phys. {\bf B276} (1986) 391 and Nucl. Phys. {\bf
B291} (1987) 876.
\bibitem{BP} D. Brecher and M.J. Perry, Nucl. Phys. {\bf B527} (1998) 121,
{\tt hep-th/9801127}; K. Behrndt, {\it Open Superstring in Non-Abelian
Gauge Field}, in the procs. of the XXIII Int. Symp. Ahrenshoop 1989, 174,
Akademie der Wissenschaften der DDR; {\it Untersuchung der Weyl-Invarianz
im Verallgemeinter $\sigma$-Modell f\"ur Offene Strings}, PhD thesis,
Humboldt-Universit\"at zu Berlin, 1990.
\bibitem{Tstr}A.A. Tseytlin, Nucl. Phys. {\bf B501} (1997) 41, {\tt
hep-th/9701125}.
\bibitem{HT} A. Hashimoto and W. Taylor, Nucl. Phys. {\bf B503} (1997)
193, {\tt hep-th/9703217}.
\bibitem{DST}  F. Denef, A. Sevrin and J. Troost, Nucl. Phys.{\bf B581}
 (2000) 135, {\tt hep-th/0002180}.
\bibitem{JT} J. Troost, 
Nucl. Phys. {\bf B568} (2000) 180, {\tt hep-th/9909187}.
\bibitem{AC} A. Abouelsaood, C. Callan, C. Nappi and S. Yost, Nucl. Phys.
{\bf B280} (1987) 599.
\bibitem{VB} P. van Baal, Comm. Math. Phys. {\bf 94} (1984) 397 and {\bf 85}
(1982) 529.
\bibitem{B} P. Bain, {\it On the Non-Abelian Born-Infeld Action}, to
appear in the proceedings of the Carg\`ese '99 Summer School,
{\tt hep-th/9909154}.
\bibitem{simon} B. Simon, {\it Representations of Finite  and Compact Groups},
Graduate Studies in Mathematics Vol.~10, American Mathematical Society 1996.\\
A. Speiser, {\it Die theorie der Gruppen van endlicher Ordnung} (4te Auflage), 
Birkh\"auser Verlag, Basel 1956.\\
D.~E.~Littlewood, {\it The theory of group characters} (2nd ed.),
Clarendon press, Oxford~1958.
\bibitem{BDL}  M.~Berkooz, M.~R.~Douglas, R.~Leigh,
Nucl. Phys. {\bf B480} (1996) 265, {\tt hep-th/9606139}.
\bibitem{brech} D. Brecher, Phys. Lett. {\bf B442} (1998) 117, {\tt 
hep-th/9804180}.
\bibitem{BDSip} E. Bergshoeff, M. de Roo and A. Sevrin, in preparation.
\bibitem{BMT} E. Bergshoeff, M. Rakowski and E. Sezgin, Phys. Lett.
{\bf B185} (1987) 371; R. Metsaev and M. Rakhmanov, Phys. Lett.
{\bf B193} (1987) 202.
\bibitem{zanon} S.~Cecotti and S.~Ferrara, Phys.~Lett.~{\bf B187} (1987)
335; S.~Ketov, {\it $N=1$ and $N=2$ Supersymmetric Nonabelian Born-Infeld
Actions from Superspace}, {\tt hep-th/0005265}; A.~Refolli, N.~Terzi and
D.~Zanon, Phys.~Lett.~{\bf B486} (2000) 337, {\tt hep-th/0006067}
\bibitem{BdS} E.A. Bergshoeff, M. de Roo and A. Sevrin, preprints
{\tt hep-th/0011018}, {\tt hep-th/0011264} and {\tt hep-th/0010151}.
\bibitem{CS} L. Cornalba, {\it On the general structure of the non-abelian 
Born-Infeld action}, preprint {\tt hep-th/0006018}.
\end{thebibliography}
\end{document}